\pdfoutput=1
\RequirePackage[T1]{fontenc}
\documentclass[12pt]{article}

\usepackage[height=8.85in,width=6.45in]{geometry}

\usepackage[utf8]{inputenc}
\usepackage{amsmath}
\usepackage{amssymb}
\usepackage{mathtools}
\numberwithin{equation}{section}
\usepackage{slashed}
\usepackage{braket}
\usepackage{enumitem}
\usepackage[svgnames,psnames]{xcolor}
\usepackage[colorlinks,citecolor=DarkGreen,linkcolor=FireBrick,urlcolor=FireBrick,linktocpage,unicode,psdextra]{hyperref}
\usepackage{cite}
\usepackage{graphicx}
\usepackage{tikz}
\usetikzlibrary{shapes.geometric}
\usetikzlibrary{calc}
\tikzstyle{circ}=[circle,draw,inner sep=1pt]
\tikzstyle{hexagon}=[draw, regular polygon,regular polygon sides=6,inner sep=2pt]
\tikzset{>=stealth}
\usepackage[bottom]{footmisc}

\usepackage{times}
\usepackage{courier}
\usepackage{bm}
\usepackage{subfig}

\usepackage{xcolor}
\usepackage{mdframed}
\newenvironment{claim}{  \begin{mdframed}[linecolor=black!0,backgroundcolor=black!10]\noindent\slshape\ignorespaces}{\end{mdframed}}

\renewenvironment{figure}[1][]{
  \begin{originalfigure}[#1]
    \begin{mdframed}[linecolor=black!0,backgroundcolor=black!1]
}{
    \end{mdframed}
  \end{originalfigure}
}

\usepackage{xcolor}

\def\Nequals#1{$\mathcal{N}{=}#1$}
\def\tr{\mathop{\mathrm{tr}}}
\def\bZ{\mathbb{Z}}
\def\bR{\mathbb{R}}

\begin{document}

\begin{titlepage}

\begin{flushright}
\end{flushright}

\vskip 3cm

\begin{center}

{\Large \bfseries  On  \boldmath\Nequals4 supersymmetry enhancements\\[.8em]
 in three dimensions}

\vskip 1cm
Benjamin Assel$^1$, 
Yuji Tachikawa$^2$ and
Alessandro Tomasiello$^3$
\vskip 1cm

\begin{tabular}{ll}
1& Paris, France. benjamin.assel@gmail.com \\
2 & Kavli Institute for the Physics and Mathematics of the Universe (WPI), \\
& University of Tokyo,  Kashiwa, Chiba 277-8583, Japan \\
& yuji.tachikawa@ipmu.jp\\
3 & Dipartimento di Matematica, Universit\`a di Milano-Bicocca, and\\
& INFN, sezione di Milano-Bicocca, I-20126 Milano, Italy\\
& alessandro.tomasiello@unimib.it
\end{tabular}

\vskip 2cm

\end{center}

\noindent 
We introduce a class of 3d theories consisting of strongly-coupled \Nequals4 systems coupled to \Nequals3 Chern--Simons gauge multiplets,
which exhibit \Nequals4 enhancements when a peculiar condition on the Chern--Simons levels is met. 
An example is  the $SU(N)^3$ Chern--Simons theory coupled to the 3d $T_N$ theory,
which enhances to \Nequals4 when $1/k_1+1/k_2+1/k_3=0$.
We also show that some but not all of these \Nequals4 enhancements can be understood by considering M5-branes on a special class of Seifert manifolds.
Our construction provides a large class of \Nequals4 theories
which have not been studied previously.

\end{titlepage}

\setcounter{tocdepth}{2}
\tableofcontents


\section{Introduction and summary}
\label{sec:introduction}

The aim of this paper is to discuss a class of three-dimensional \Nequals4 supersymmetric theories 
which have rarely been studied in the literature.
There are two complementary motivations leading to the same class of theories, 
as we will see below.

Our first motivation is in the context of highly-supersymmetric Chern--Simons-matter theories.
As is well-known, the highest supersymmetry one can achieve in generic Chern--Simons-matter systems is  \Nequals3 \cite{Schwarz:2004yj,Gaiotto:2007qi}.
It is also known that, when the matter content and the gauge group are carefully chosen,
the supersymmetry can show further enhancements to \Nequals4 \cite{Gaiotto:2008sd,Hosomichi:2008jd}, 
\Nequals5 \cite{Hosomichi:2008jb}, \Nequals6 \cite{Aharony:2008ug,Aharony:2008gk},
and \Nequals8 \cite{Gustavsson:2007vu,Bagger:2007jr}.
Almost all of these theories use only bifundamental matter fields, except 
a single case mentioned implicitly in \cite{Gaiotto:2008sd},
which is an $SU(2)^3$ Chern--Simons (CS) theory coupled to a half-hypermultiplet in the trifundamental representation.\footnote{%
In 3d supersymmetric Chern--Simons theories, there is a different  mechanism of supersymmetry enhancement,
via the emergence of additional supersymmetry generators in the monopole sector,
for example the ABJM theory at level $1$ or $2$ \cite{Bashkirov:2010kz}.
In our case, the supersymmetry enhancement happens in the non-monopole sector.}

This theory is generically \Nequals3 supersymmetric, but shows \Nequals4 enhancement 
if and only if the three levels $k_{1,2,3}$ satisfy the relation $1/k_1 + 1/k_2 + 1/k_3=0$.
What we do here is to generalize this example using another idea already mentioned in \cite{Gaiotto:2008sd}, 
of using strongly-coupled \Nequals4 systems with curved Higgs branches as the matter content.
For example, we will show that the theory composed of $SU(N)^3$ Chern--Simons vector multiplets coupled to the 3d $T_N$ theory enjoys \Nequals4 enhancement in the same manner, if and only if $1/k_1+1/k_2+1/k_3=0$.
We will further generalize it to a class-S-like construction.\footnote{%
Generalization to the $T_{G}$ theory gauged by $G^3$ for simply-laced $G$ is immediate, 
although we only discuss the case of $G=SU(N)$ in the main text for brevity.
}

Our second motivation is in the study of supersymmetric compactifications of M5-branes on 3-manifolds.
Their general study was initiated in \cite{Terashima:2011qi} for mapping tori and \cite{Dimofte:2011ju} for triangulated manifolds.
In these cases, the 3-manifolds are hyperbolic, and
the resulting theories generically have \Nequals2 supersymmetry.
A far simpler class of theories was studied slightly before them in \cite{Benini:2010uu},
where the compactifications on $S^1$ times surfaces were considered.
They can be studied as the $S^1$ compactification of 4d class S theories,
and have \Nequals4 supersymmetry.

In the general classification of 3-manifolds conjectured by Thurston and proved recently,
there are manifolds which lie between these two extremes, known as
Seifert manifolds and graph manifolds.
Such compactifications have not been studied as thoroughly as the hyperbolic ones,
but the 3d theories which result from them have already been worked out 
in e.g.~\cite{Gadde:2013sca,Pei:2015jsa,Gukov:2016gkn,Gukov:2017kmk,Eckhard:2019jgg,Cho:2020ljj},
although determining them was usually not the main objective of these papers.
Generically they have \Nequals2 supersymmetry.
In this paper, we find several spaces which lead to \Nequals4 supersymmetry. 
A simple subclass consists of Seifert fibrations over a sphere with three singular fibers, whose Seifert parameters $q_i/p_i$ for $i=1,2,3$ sum to zero, $\sum_i q_i/p_i=0$.
When $q_{1,2,3}=1$, the resulting 3d theories turn out to be equivalent to 
the theories we mentioned above, coming from our first motivation.

In this particular subclass, the supersymmetry enhancement we observe is the same
both in the field theoretical analysis and in the analysis using the geometry of 3-manifolds.
This, however, is not always the case,
and there are a large number of examples where 
the supersymmetry seen field theoretically is higher than what can be gleaned from the geometry, 
as we will describe in more detail in the rest of the paper.
The authors would hope to come back to this issue in the future.

The rest of the paper is organized as follows.
In Sec.~\ref{sec:highly}, we start by recalling the supersymmetry enhancement mechanism in 3d Chern--Simons-matter systems.
We then discuss the enhancement of the supersymmetry  from \Nequals3 to \Nequals4 of $SU(2)^3$ Chern--Simons theory coupled to a half-hypermultiplet in the trifundamental representation,
and study how it can be generalized to the case of $SU(N)^3$ Chern--Simons theory coupled to the $T_N$ theory.

After making some further generalization, we move on to Sec.~\ref{sec:3},
where we consider M5-branes compactified on 3-manifolds.
After very briefly reviewing the classification of 3-manifolds,
we study M5-branes on Seifert manifolds and on graph manifolds.
We will see that the resulting 3d theories realize the 3d Chern--Simons-matter systems we discuss in Sec.~\ref{sec:highly}.
We will also see that the reduction of the holonomy of 3-manifolds used in the construction
does explain the \Nequals4 enhancement in some cases, but not all.

In Appendix~\ref{sec:homology} and in Appendix~\ref{sec:sugra}, 
we study the homology groups and the possible supergravity backgrounds on the 3-manifolds used in Sec.~\ref{sec:3},
in order to find any indication of supersymmetry enhancement when the holonomy reduction does not happen.
Unfortunately, we do not find any definitive results, although we do find some hints.

In the final Appendix~\ref{app:wall}, we make some further comments 
on the property of the 3d $SL(2,\bZ)$ duality wall theory,
and of the theory obtained by diagonally gauging its $SU(N)$ flavor symmetry.
This is done using the results obtained in the course of Sec.~\ref{sec:3}
and the computation of the contact terms using localization, following \cite{Closset:2012vg,Closset:2012vp}.

Before proceeding, we pause here to mention that
we will be cavalier about the choice of the global structure of the gauge group and also about the possible presence of almost decoupled topological sectors. 
In particular, we will stick to the careless habit of not distinguishing Lie algebras and Lie groups in the notations, which was prevalent in our community until several years ago.
It would surely be interesting to study these issues carefully,
which was in fact one of the main motivations in the previous study \cite{Eckhard:2019jgg} on 3d theories coming from Seifert manifolds, although not in the context of supersymmetry enhancements.

We also note that there recently appeared a paper \cite{Choi:2022dju} 
where, among others, \Nequals4 enhancements of 6d \Nequals{(2,0)} theories on non-hyperbolic 3-manifolds $M$ were discussed. 
The results here and there cannot be readily compared,  
because their derivation is rather different, and also because they considered 3d theories $T_\text{irred}[M]$ associated to irreducible $SL(2)$ connections on $M$,
while we consider $T_\text{full}[M]$ which is the ordinary compactification on $M$; see \cite{Chung:2014qpa,Gang:2018wek} for the difference.
That said, it would be interesting to study the interconnection between the two approaches,
as the 3-manifolds discussed in the two papers do overlap.

\section{Highly-supersymmetric Chern--Simons-matter theories}
\label{sec:highly}

In this section, we introduce the readers to our first class of 3d Chern--Simons-matter theories
which show enhancements to \Nequals4 supersymmetry.
This is done from the study of highly-supersymmetric Chern--Simons-matter theories.

When the matter content is generic, supersymmetric Chern--Simons-matter theories are
only possible up to \Nequals3 supersymmetry.
Theories with higher supersymmetry were first constructed 
in \cite{Gustavsson:2007vu,Bagger:2007jr} in the \Nequals8 cases using 3-algebras.
The construction was motivated by the then-popular beliefs that something other than ordinary Lie algebras would be necessary.

It was soon realized, however, that they can be written in terms of ordinary Lie algebras \cite{Bandres:2008vf},
and generalizations to \Nequals4 theories were performed using the 3d \Nequals1 superfield formalism in \cite{Gaiotto:2008sd,Hosomichi:2008jd}.
Slightly later, the famous \Nequals6 theory of Aharony, Bergman, Jafferis and Maldacena was found in \cite{Aharony:2008ug} using the 3d \Nequals2 superfield formalism.
In fact, all the supersymmetry enhancements listed here can be understood 
by adapting this 3d \Nequals2 superfield method \cite{Schnabl:2008wj}.
Here we provide a quick review of the mechanism of the enhancement,
and then introduce our first class of theories showing \Nequals4 enhancements which were not studied previously.

\subsection{Mechanism of the enhancement}
\paragraph{Enhancement from \Nequals2 to \Nequals4 in 4d:}
Let us start by recalling a method to understand the structure of 4d \Nequals4 super Yang--Mills theories.
It is well-known that we can write down a 4d \Nequals2 gauge theory for any gauge group and hypermultiplets. 
Let us pick a gauge group $G$ and let the hypermultiplet be in its adjoint representation.
In the \Nequals1 language, the superpotential has the form \begin{equation}
W \propto \tr \Phi[A,B]
\end{equation}
where $\Phi$ is the scalar in the vector multiplet and $(A,B)$ form the hypermultiplet.
In the \Nequals1 perspective, it is clear that the theory is symmetric 
under the $SU(3)_F$ flavor symmetry acting on three adjoint scalars $\Phi$, $A$ and $B$,
from the structure of the superpotential.
Meanwhile, in the \Nequals2 perspective, the $SU(2)_R$ symmetry acts on $(A, B^\dagger)$.
Therefore, $SU(2)_R$ and $SU(3)_F$ do not commute, and necessarily combine to a larger R-symmetry, 
signaling the enhancement of the supersymmetry to \Nequals4.

\paragraph{Enhancement from \Nequals3 to \Nequals{N} in 3d:}
In three dimensions, we can similarly achieve \Nequals3 supersymmetry with general gauge group $G=\prod_i G_i$ with non-zero Chern--Simons terms, and an arbitrary representation $R$ for the hypermultiplets.
The superpotential in \Nequals2 language is \begin{equation}
W \propto \sum_i \left(\tr \Phi_i \mu_i - \frac{k_i}2 \tr \Phi_i^2\right),\label{N=4W}
\end{equation}
where  $\Phi_i$ is the adjoint scalar in the $i$-th Chern--Simons supermultiplet 
of level $k_i$
and $\mu_i$ is the moment map operator constructed from the hypermultiplets.

We note that this system preserves \Nequals4 supersymmetry when $k_i=0$.
In this case, the superpotential possesses a manifest $U(1)$ symmetry assigning charge $+2$ and $-2$ to $\mu_i$ and $\Phi_i$.
This is in fact a part of $SO(4)_R$ symmetry of the \Nequals4 supersymmetry algebra,
and is broken by the term proportional to $k_i$ in \eqref{N=4W} when it is nonzero.

For the moment, let us assume $k_i\neq 0$.
As the adjoint scalars $\Phi_i$ have no kinetic term, they can be integrated out, giving 
\begin{equation}
W \propto\sum_i \frac1{k_i}\tr \mu_i^2.\label{N=2W}
\end{equation}
When the matter representations and the levels are chosen carefully, this $W$ can have a flavor symmetry not commuting with the \Nequals3 R-symmetry,
signaling the enhancement to supersymmetry higher than \Nequals3.

In three dimensions, an \Nequals{N} superconformal theory has $SO(N)_R$ flavor symmetry.
The \Nequals2 formalism makes only $SO(2)_R$ symmetry manifest,
and therefore we should see an additional $SO(N-2)$ flavor symmetry 
in the \Nequals2 superpotential \eqref{N=2W} to have the enhancement. 

\paragraph{\Nequals6 theories of ABJ(M):}
As an example, consider the \Nequals3 $U(N)_k\times U(N')_{-k'}$ Chern--Simons theory with two bifundamentals $A_i$, $B^i$ where $i=1,2$.
The superpotential \eqref{N=2W} \begin{equation}
W\propto \frac1k \tr (A_i B^i)^2 - \frac1{k'} \tr (B^iA_i)^2 
\end{equation} simplifies, if $k=k'$, to \begin{equation}
\propto \frac1k \tr A_i B^a A_j B^b \epsilon^{ij}\epsilon_{ab}
\end{equation} showing the flavor symmetry $SU(2)\times SU(2)=SO(6-2)$.
This flavor symmetry does not commute with $SO(3)_R$, under which $A_i$ forms a doublet with $(B^i)^\dagger$.
This means that the theory enhances to \Nequals6 when $k=k'$;
this is the famous theory of Aharony-Maldacena-Bergman-Jafferis or Aharony-Bergman-Jafferis, 
depending on $N=N'$ or $N\neq N'$ \cite{Aharony:2008ug,Aharony:2008gk}.

\paragraph{\Nequals4 theories of Gaiotto--Witten:}
As another class of examples, consider the \Nequals3 theory with a general gauge group  $G=\prod_i G_i$ and (half-)hypermultiplets in a general representation $R$.
Let us suppose that the superpotential \eqref{N=2W} simply vanishes, \begin{equation}
W \propto\sum_i \frac1{k_i}\tr \mu_i^2=0,\label{fundamental}
\end{equation}
thanks to a careful choice of $G$ and $R$.

Let us see that this means that the theory enhances to \Nequals4 supersymmetry.
Before coupling to \Nequals3 Chern--Simons multiplets, 
the theory of free (half-)hypermultiplets has \Nequals4 supersymmetry, with four supercharges
$Q_{1,2,3,4}$, with R-symmetries $J_{ij}$ mapping $Q_i$ to $Q_j$.

Let us say that we are using an \Nequals2 subalgebra including $Q_{1,2}$ and $J_{12}$ 
for our \Nequals2 description.
When the superpotential vanishes as in \eqref{fundamental},
we have a $U(1)=SO(4-2)$ flavor symmetry, assigning the charge $+1$ for all \Nequals2 chiral multiplets. 
This $U(1)$ symmetry is $J_{34}$ of the original theory of hypermultiplets.

Now, the \Nequals3 Chern--Simons multiplets are known to preserve $Q_{1,2,3}$.
We now found that $J_{34}$ is preserved. Therefore, $Q_4$ is also preserved,
meaning that the gauged theory has \Nequals4 supersymmetry.
The condition was originally obtained by Gaiotto and Witten in \cite{Gaiotto:2008sd},
using the \Nequals1 superfield formalism.

The construction can be generalized by replacing (half-)hypermultiplets by a more general strongly-coupled 3d \Nequals4 theory $T$ with $G$ symmetry with $SO(4)_R$ symmetry.
We note that such a theory $T$, regarded as an \Nequals2 theory, 
has a $U(1)$ flavor symmetry $J_{34}$ assigning charge $+2$ to the moment map operators $\mu$.
When we gauge the $G$ symmetry with \Nequals3 Chern--Simons couplings,
the \Nequals2 superpotential \eqref{N=2W} breaks this $U(1)$ flavor symmetry for generic levels.
However, when the relation \eqref{fundamental} is satisfied, 
this $U(1)=SO(4-2)$ flavor symmetry is restored,\footnote{%
Note that this symmetry is not restored 
if the relation \eqref{fundamental} holds only as a chiral ring relation
but not as an actual operator equation,
since in that case the integral $\int d^2\theta W$ can still be nonzero and violate the $U(1)$ symmetry $J_{34}$.
Luckily for us, when $T$ is superconformal, the relation \eqref{fundamental} in the chiral ring automatically implies the validity at the level of operators,
since the moment map operators $\mu_i$ are chiral primaries,
and therefore $\sum \tr (\mu_i)^2 /k_i$ is also a chiral primary.
See also the discussions around the end of \cite[Sec.~3.2.2]{Gaiotto:2008sd}.
}
and the gauged theory enhances to \Nequals4.
This strongly-coupled generalization was already mentioned in \cite{Gaiotto:2008sd},
although in the \Nequals1 superfield language.

It was also found in the original article \cite{Gaiotto:2008sd} that 
the condition \eqref{fundamental}
when the matter content is given by (half-)hypermultiplets
is equivalent to the statement that $\mathfrak{g}\oplus R$ forms a super Lie algebra with an invariant non-degenerate trace.
In more detail,
1) $\mathfrak{g}$ and $R$ are the bosonic and the fermionic part of the super Lie algebra, respectively,
2) the vanishing \eqref{fundamental} is equivalent to the fermion-fermion-fermion part of the super Jacobi identity,
and 3) the ratio of the levels $\{k_i\}$ is a part of the structure constants.\footnote{%
This somewhat unexpected appearance of a super Lie algebra can be made
less mysterious by noting that the Rozansky-Witten topological twist of the
resulting \Nequals4 theory
is the Chern--Simons theory based on the super Lie algebra $\mathfrak{g}\oplus R$, as was found in \cite{Kapustin:2009cd}.}
As the classification of super Lie algebras with invariant traces is long known \cite{Kac},
this gives us all theories where Chern--Simons gauge fields couple to (half-)hypermultiplets 
such that the supersymmetry enhances to \Nequals4 from the condition \eqref{fundamental}.

An example of this type of theories is to take the super Lie algebra to be $OSp(2m|2n)$, 
which corresponds to the gauge theory $O(2m)_{2k} \times USp(2n)_{-k}$ 
with a half-hypermultiplet in the bifundamental representation. 
This turns out to be the starting point of our generalization.

\subsection{New \Nequals{4} theories}
\subsubsection{Gauged $T_N$ theory}
\label{sec:mainexamples}

Let us now consider the special case of $OSp(4|2)$.
The corresponding gauge theory has the gauge group $SO(4)\times USp(2) \simeq SU(2)_1\times SU(2)_2\times SU(2)_3$,
with a half-hypermultiplet in $\mathbf{4}\otimes \mathbf{2} \simeq \mathbf{2}_1\otimes \mathbf{2}_2\otimes \mathbf{2}_3$,
which we denote by $Q_{aiu}$.
The condition for the \Nequals4 enhancement \eqref{fundamental} is \begin{equation}
W \propto \frac{1}{k_1} \tr (\mu_1)^2 
+\frac{1}{k_2} \tr (\mu_2)^2 
+\frac{1}{k_3} \tr (\mu_3)^2  = 0
\label{xxx}
\end{equation}
where $k_{1,2,3}$ and $\mu_{1,2,3}$ are the levels and the moment map operators for $SU(2)_{1,2,3}$.
More explicitly, $\mu_{1,2,3}$ are given by 
{\let\green\relax\let\orange\relax\let\alert\relax
\begin{equation}
\mu_{1}{}^{\green{c}}{}_{\green{b}}= \epsilon^{\green{ca}}\mu_{1,\green{ab}},\quad
\mu_{2}{}^{\orange{k}}{}_{\orange{j}}= \epsilon^{\orange{ki}} \mu_{2,\orange{ij}},\quad
\mu_{3}{}^{\alert{w}}{}_{\alert{v}}= \epsilon^{\alert{wu}} \mu_{3,\alert{uv}}
\end{equation}
where
\begin{equation}
\mu_{1,\green{ab}} =\epsilon^{\orange{ij}}\epsilon^{\alert{uv}} Q_{\green{a}\orange{i}\alert{u}}Q_{\green{b}\orange{j}\alert{v}},\quad
\mu_{2,\orange{ij}}=\epsilon^{\green{ab}}\epsilon^{\alert{uv}} Q_{\green{a}\orange{i}\alert{u}}Q_{\green{b}\orange{j}\alert{v}},\quad
\mu_{3,\alert{uv}}=\epsilon^{\green{ab}}\epsilon^{\orange{ij}} Q_{\green{a}\orange{i}\alert{u}}Q_{\green{b}\orange{j}\alert{v}}
\end{equation}}\relax which satisfy the crucial relation\footnote{%
This quantity is known as Cayley's hyperdeterminant of $Q_{aiu}$.
} \begin{equation}
\tr(\mu_1)^2=\tr(\mu_2)^2=\tr(\mu_3)^2.\label{crucial}
\end{equation} 
Therefore, the enhancement condition \eqref{xxx} is satisfied if and only if we have \begin{equation}
\frac{1}{k_1}+\frac{1}{k_2}+\frac{1}{k_3}=0.\label{yyy}
\end{equation}
Physically, we need to require that the levels are integers.\footnote{%
This is when the global structure of the gauge group is $SU(2)^3$.
The structure of $Q_{aiu}$ allows one to take the gauge group to be $SU(2)^3/C$
where $C\simeq (\bZ_2)^2\subset \{\pm1\}^3\subset SU(2)^3$ is the subgroup of the center trivially acting on $Q_{aiu}$.
Then the Chern--Simons levels will be slightly more constrained.
We will not discuss similar complications coming from the global structure of the gauge group any further in this paper.
}
Luckily for us, the relation \eqref{yyy} has many integer solutions, e.g.~$(k_1,k_2,k_3)=(pq,qr,rp)$ with $p+q+r=0$ and $p,q,r\in\bZ_{\neq 0}$.\footnote{%
These in fact exhaust all solutions up to rescalings. To see this,
write $(k_1,k_2,k_3)=k(a,b,c)$ with $\gcd(a,b,c)=1$.
By changing the sign of $k$, we can set $a+c\le 0$ without loss of generality.
The requirement \eqref{yyy} can now be rewritten as $(a+c)(b+c)=c^2$.
Now, $a+c$ and $b+c$ are coprime, since if not, 
let a prime $l$ divides both $x:=a+c$ and $y:=b+c$. 
Then $l$ also divides $c$, contradicting $\gcd(a,b,c)=1$.
Now, the product of two coprime integers $xy$ being a square $c^2$
means that $x$, $y$ are themselves squares times $\pm1$.
As we assumed $x\le 0$,
we see that there are two coprime integers $p$ and $r$ such that 
$x=-p^2$, $y=-r^2$ and $c=rp$,
and therefore $(a,b,c)=(pq,qr,rp)$ where $q=-p-r$.
}

As mentioned above, the ratio of the levels solving \eqref{xxx} translates to the structure constants of the super Lie algebra.
The relation \eqref{yyy} means that $OSp(4|2)$ has a one-parameter family of structure constants.
In fact, among the classification in \cite{Kac} of super Lie algebras, 
$D(2,1)=OSp(4|2)$ is the only case where the structure constants come in a family, 
and is often denoted as $D(2,1;\alpha)$ to make this fact explicit.

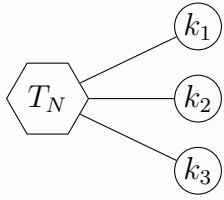
\begin{figure}
\centering
\begin{tikzpicture}[yscale=.8]
\node[hexagon] (TN) at (0,0) {$T_N$};
\node[circ] (k1) at (2,1.2) {$k_1$};
\node[circ] (k2) at (2,0) {$k_2$};
\node[circ] (k3) at (2,-1.2) {$k_3$};
\draw (TN)--(k1);
\draw (TN)--(k2);
\draw (TN)--(k3);
\end{tikzpicture}
\caption{$T_N$ theory gauged with $SU(N)_{k_1}\times SU(N)_{k_2}\times SU(N)_{k_3}$. 
The supersymmetry enhances to \Nequals4 when $1/k_1+1/k_2+1/k_3=0$. \label{fig:gaugedTn}}
\end{figure}

Let us now generalize this example, by replacing $SU(2)^3$ and $Q_{aiu}$
by $SU(N)^3$ and the 3d $T_N$ theory.
Here the 3d $T_N$ theory is a strongly-coupled 3d \Nequals4 SCFT
having $SU(N)^3$ symmetry,
obtained by compactifying the 4d $T_N$ theory on $S^1$.
Equivalently, it is given by wrapping $N$ M5-branes on $S^1$ times a sphere with three full punctures.
Its chiral ring is summarized e.g.~in \cite{Tachikawa:2015bga}.
Crucially, the three $SU(N)$ moment map operators $\mu_{1,2,3}$ satisfy the relations \eqref{crucial}, as we will see in Sec.~\ref{sec:ingredients};
therefore the combined system has enhanced \Nequals4 supersymmetry
if the relation \eqref{yyy} is satisfied. 
The structure of the theory can be depicted as in Fig.~\ref{fig:gaugedTn}.

There is an easy generalization of this construction if one knows the basics of 3d class S theories,
first studied in \cite{Benini:2010uu}.
Let us take two copies of the 3d $T_N$ theory, and gauge a diagonal subgroup of two $SU(N)$ flavor symmetries, each belonging to a separate $T_N$ theory, by an \Nequals4 $SU(N)$ vector multiplet.
This is the compactification of 6d \Nequals{(2,0)} theory on $S^1$ times a sphere with four punctures.
The resulting theory has $SU(N)^4$ flavor symmetry, whose moment map fields satisfy \begin{equation}
\tr(\mu_1)^2=\tr(\mu_2)^2=\tr(\mu_3)^2 = \tr(\mu_4)^2.
\end{equation}
More generally, we can take $n-1$ copies of the $T_N$ theory and couple \Nequals4 $SU(N)$ vector multiplets appropriately,
and realize the compactification of 6d \Nequals{(2,0)} theory on $S^1$ times a sphere with $n$  punctures.
Let us denote the resulting theory by $T_{N,n}$.
It has $SU(N)^n$ symmetry, whose moment map fields satisfy the basic relation that \begin{equation}
\tr(\mu_i)^2 \quad \text{is independent of $i=1,2,\ldots, n$.}
\label{superimportant}
\end{equation}
The derivation of this important relation is reviewed in Sec.~\ref{sec:ingredients}.

We can then couple the $SU(N)^n$ flavor symmetry to \Nequals3 $SU(N)$ Chern--Simons gauge fields
with levels $k_i$ for $i=1,\ldots, n$.
Exactly as before, we find that the superpotential after the elimination of adjoint scalars 
vanishes when \begin{equation}
\sum_i \frac1{k_i}=0,
\end{equation}
leading to  \Nequals4 enhancements.

So far these theories might look like a mere curiosity. 
In the next section we are going to find a geometric interpretation for this enhancement.
Before getting there, let us consider a generalization of this construction.

\subsubsection{Further generalizations}
\label{sec:generalizations}
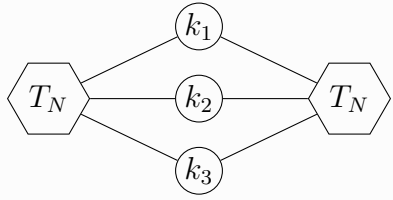
\begin{figure}
\centering
\begin{tikzpicture}[yscale=.8]
\node[hexagon] (TN) at (0,0) {$T_N$};
\node[hexagon] (TN') at (4,0) {$T_N$};
\node[circ] (k1) at (2,1.2) {$k_1$};
\node[circ] (k2) at (2,0) {$k_2$};
\node[circ] (k3) at (2,-1.2) {$k_3$};
\draw (TN)--(k1)--(TN');
\draw (TN)--(k2)--(TN');
\draw (TN)--(k3)--(TN');
\end{tikzpicture}
\caption{Two $T_N$ theories gauged with $SU(N)_{k_1}\times SU(N)_{k_2}\times SU(N)_{k_3}$. \label{fig:genus2}}
\end{figure}

Let us now consider the theory depicted in Fig.~\ref{fig:genus2}.
Namely, we take two copies of 3d $T_N$ theory.
For each $SU(N)$ symmetry of one $T_N$ theory,
we pick an $SU(N)$ symmetry from the other $T_N$ theory,
and couple their diagonal combination to an \Nequals3 $SU(N)$ Chern--Simons multiplet.
Denoting the levels by $k_{1,2,3}$, we see that the superpotential in the \Nequals2 description is
\begin{equation}
\begin{aligned}
W= &-\frac{k_1}2 \tr (\Phi_1)^2 + \tr \Phi_1 (\mu_1^{(1)}+\mu_1^{(2)})\\
&-\frac{k_2}2 \tr (\Phi_2)^2 + \tr \Phi_2 (\mu_2^{(1)}+\mu_2^{(2)})\\
&-\frac{k_3}2 \tr (\Phi_3)^2 + \tr \Phi_3 (\mu_3^{(1)}+\mu_3^{(2)})
\end{aligned}
\end{equation}
where $\Phi_i$ is the adjoint scalar in the $i$-th \Nequals3 Chern--Simons multiplet
and $\mu_{1,2,3}^{(a)}$ is the moment map field of $SU(N)^3$ of the $a$-th $T_N$ theory.

Eliminating $\Phi_{1,2,3}$, we find that the superpotential becomes \begin{equation}
\begin{aligned}
W =& \frac12\left(\frac{1}{k_1}+\frac{1}{k_2}+\frac{1}{k_3}\right) (\tr (\mu^{(1)})^2 + \tr (\mu^{(2)})^2) \\
& +\frac1{k_1} \tr \mu_1^{(1)}\mu_1^{(2)}
+\frac1{k_2} \tr \mu_2^{(1)}\mu_2^{(2)}
+\frac1{k_3} \tr \mu_3^{(1)}\mu_3^{(2)}
\end{aligned}
\end{equation}
where we used the chiral ring relation 
$\tr (\mu_1^{(1)})^2 =\tr (\mu_2^{(1)})^2=\tr (\mu_3^{(1)})^2=:\tr (\mu^{(1)})^2$
for the first $T_N$ theory, and similarly for the second copy.
The superpotential simplifies when \begin{equation}
\frac{1}{k_1}+\frac{1}{k_2}+\frac{1}{k_3}=0
\end{equation} to \begin{equation}
W= \frac1{k_1} \tr \mu_1^{(1)}\mu_1^{(2)}
+\frac1{k_2} \tr \mu_2^{(1)}\mu_2^{(2)}
+\frac1{k_3} \tr \mu_3^{(1)}\mu_3^{(2)},\label{ZZZ}
\end{equation}
which has a $U(1)=SO(4-2)$ flavor symmetry assigning charge $+2$ to the moment map fields of the first $T_N$ theory and charge $-2$ to those of the second $T_N$ theory.
This signifies that the theory enhances to \Nequals4.

To give some more detail, let $Q^{(a)}_{1,2,3,4}$ be four supercharges of the $a$-th $T_N$ theory,
and $J^{(a)}_{ij}$ be the R-symmetry of the $a$-th theory sending $Q^{(a)}_i$ to $Q^{(a)}_j$.
The \Nequals3 Chern--Simons couplings preserve $Q_i:=Q^{(1)}_i+Q^{(2)}_i$ for $i=1,2,3$,
and our \Nequals2 formalism makes $Q_1$, $Q_2$ and $J_{12}:=J_{12}^{(1)}+J_{12}^{(2)}$ manifest.
The $U(1)$ symmetry preserved by the simplified superpotential \eqref{ZZZ} is 
$J_{34}':=J_{34}^{(1)}-J_{34}^{(2)}$.
Note the relative minus sign, assigning opposite $U(1)$ charges to $\mu_i^{(1)}$ and $\mu_i^{(2)}$.
This $U(1)$ symmetry sends $Q_3=Q^{(1)}_3+Q^{(2)}_3$ to
$Q_4':=Q^{(1)}_4-Q^{(2)}_4$.
This type of \Nequals4 enhancements was first found in \cite{Hosomichi:2008jd} in the Lagrangian case.

Note that the operation $(Q_1,Q_2,Q_3,Q_4) \mapsto (Q_1,Q_2,Q_3,-Q_4)$ 
in an \Nequals4 theory is the 3d mirror symmetry which exchanges the Higgs branch and the Coulomb branch
and sends hypermultiplets to twisted hypermultiplets.
Therefore, when the supersymmetry enhances to \Nequals4 in this theory,
the assignment of \Nequals4 supercharges is different between the first copy 
and the second copy of the $T_N$ theory.
The $SU(2)$ R-symmetry acting on the Coulomb branch of one $T_N$ theory is mapped to the $SU(2)$ R-symmetry 
acting on the Higgs branch of the other $T_N$ theory.
In other words, there is no standard assignment of the Coulomb branch and the Higgs branch to 
the final \Nequals4 theory. 

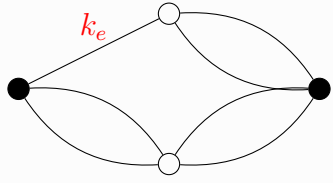
\begin{figure}
\centering
\begin{tikzpicture}
\node[circ,inner sep=1mm,fill] (B1) at (0,0) {};
\node[circ,inner sep=1mm,fill] (B2) at (4,0) {};
\node[circ,inner sep=1mm] (W1) at (2,1) {};
\node[circ,inner sep=1mm] (W2) at (2,-1) {};
\draw (B1)--node[above,red]{$k_e$} (W1);
\draw (B1) edge[bend left] (W2);
\draw (B1) edge[bend right] (W2);
\draw (B2) edge[bend left] (W2);
\draw (B2) edge[bend right] (W2);
\draw (B2) edge[bend left] (W1);
\draw (B2) edge[bend right] (W1);
\end{tikzpicture}

\caption{A graph encoding the coupling of $T_{N,n}$ theories and \Nequals3 Chern--Simons multiplets.
Every link $e$ represents a coupling to a CS level $k_e$ as in \eqref{eq:ke}.
\label{fig:general}}
\end{figure}

Let us conclude this section by making a further generalization.
We take a bipartite graph, an example of which is shown in Fig.~\ref{fig:general}.
There, each $n$-valent node is a copy of the $T_{N,n}$ theory,
obtained by compactifying 6d \Nequals{(2,0)} theory on $S^1$ times $S^2$ with $n$ full punctures.
Recall that it has $SU(N)^n$ symmetry whose moment map fields satisfy \eqref{superimportant}.

Next, each edge $e$ connecting two nodes is an \Nequals3 $SU(N)$ Chern--Simons multiplet
coupling to the diagonal subgroup of two $SU(N)$ symmetries,
one from the $T_{N,n}$ theory on a black node,
and another from the $T_{N,n}$ theory on a white node,
assigned with a non-zero Chern--Simons level $k_e$.\footnote{We assume $k_e\neq 0$ without loss of 
generality, as connecting a $T_{N,n_1}$ and a $T_{N,n_2}$ theory with an edge with zero CS level is 
simply constructing the $T_{N,n_1+n_2}$ theory, so this would be represented by a single vertex and 
no edge in the graph.}
The superpotential is given by
\begin{equation}
\label{eq:ke}
  W = \sum_{e:\text{edge}}\left[ -\frac{k_e}{2} \tr(\Phi_e)^2  + \tr \Phi_e(\mu_e^\text{black}+\mu_e^\text{white})\right].
\end{equation}

Integrating out the \Nequals{2} adjoint scalars, we get
\begin{align}
  W &= \sum_{e:\text{edge}}\left[ \frac{1}{2k_e}\big(\tr(\mu_e^\text{black})^2 + \tr(\mu_e^\text{white})^2))
   + \frac{1}{k_e}\tr\mu_e^\text{black} \mu_e^\text{white}\right] \\
   &=
   \sum_{e:\text{edge}} \frac{1}{k_e}\tr\mu_e^\text{black} \mu_e^\text{white}
   +\sum_{v:\text{node}} \sum_{\substack{e:\text{edge} \\ \text{connecting to $v$}}}
   \frac{1}{2k_e} \tr (\mu_{v,e})^2
\end{align}
where $\mu_{v,e}$ is the moment map of the $SU(N)$ flavor symmetry of the $T_{N,n}$ theory
at the node $v$ coupled to the Chern--Simons gauge multiplet for the edge $e$.
Thanks to the chiral ring relation \eqref{superimportant},
$\tr(\mu_{v,e})^2$ is independent of $e$.
Therefore, when the sum of the inverse of the three Chern--Simons levels at each node vanishes, i.e.~when
\begin{equation}
\label{eq:bipartite_k_rel}
   \sum_{\substack{e:\text{edge}\\ \text{connecting to $v$}}} \frac{1}{k_e} = 0\, \quad\text{for each node $v$},
\end{equation}
the superpotential becomes \begin{equation}
W= \sum_{e:\text{edge}} \frac{1}{k_e}\tr\mu_e^\text{black} \mu_e^\text{white}
\end{equation}
where $\mu_e^\text{black,white}$ is the $SU(N)$ moment map fields of the two $T_{N,n}$ theories
on the black node and the white node connected by the edge $e$.
This preserves the $U(1)$ symmetry \begin{equation}
J_{34} := \sum_{b: \text{black nodes}} J_{34}^{(b)} -
\sum_{w: \text{white nodes}} J_{34}^{(w)}
\end{equation} and therefore the theory has the fourth supercharge \begin{equation}
Q_4 := \sum_{b: \text{black nodes}} Q_{4}^{(b)} -
\sum_{w: \text{white nodes}} Q_{4}^{(w)}.
\end{equation}
To summarize, an \Nequals{3} theory described by a bipartite graph has its supersymmetry 
enhanced to \Nequals{4} when \eqref{eq:bipartite_k_rel} is satisfied.

\section{M5-branes on 3-manifolds}
\label{sec:3}

In this section, we consider M5-branes\footnote{%
Below, we neglect the contribution from the center-of-mass modes of the M5-branes,
which only give rise to almost decoupled free fields and/or topological degrees of freedom.
Our analysis can also be generalized to 6d \Nequals{(2,0)} theory of arbitrary type $G=A_{N-1}, D_N, E_{6,7,8}$ without any change.
For brevity, we use a somewhat imprecise language and collectively call them `M5-branes on 3-manifolds'.
} on a class of 3-manifolds and will be led to the same class of 3d theories
showing \Nequals4 enhancements which we saw in the previous section.
Our study in this section will also show us how to generalize these examples further.

It is well-known that studies of M5-branes on 3-manifolds, started in \cite{Terashima:2011qi,Dimofte:2011ju}, have uncovered various interesting properties in hyperbolic geometry, knot theory,
and 3-dimensional supersymmetric theories.
To approach this vast area of research systematically,
we find it  useful to recall the classification of 3-manifolds.

\subsection{Classification of 3-manifolds}

It is well known that surfaces are topologically classified by their genera,
and any surface can be decomposed into a number of three-punctured spheres 
by cutting along non-intersecting non-contractible circles. 
An analogous classification of 3-manifolds has been obtained by a combined efforts of
many mathematicians.
Our aim in this section is to give an extremely brief overview of this important result;
for more details, see e.g.~\cite{Scott,HatcherOnline}.

\begin{figure}
\centering
\includegraphics[width=.5\textwidth]{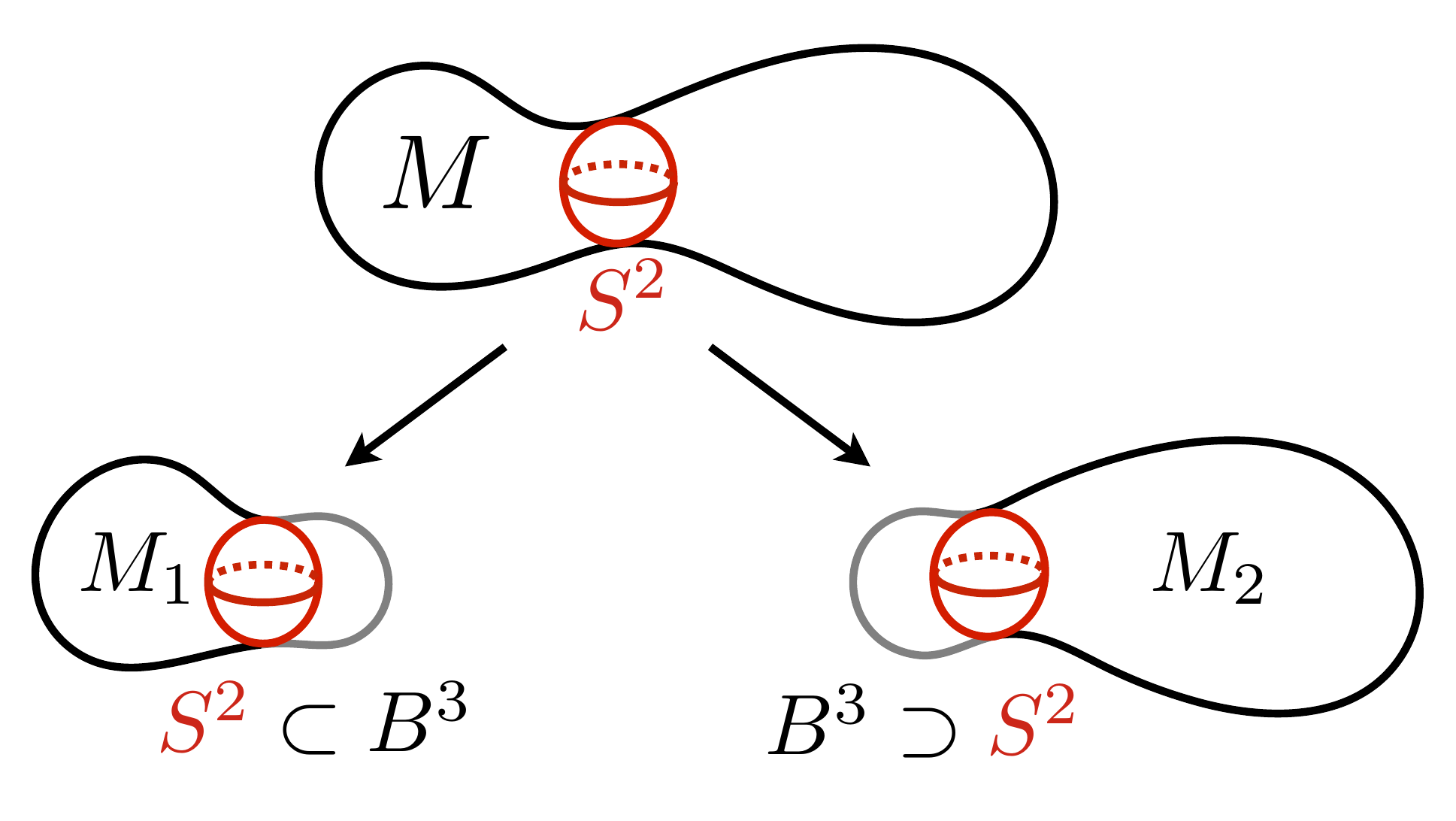}
\caption{Cutting 3-manifolds along $S^2$. \label{fig:prime}}
\end{figure}

\paragraph{Prime decomposition:}
Given two closed 3-manifolds $M_1$ and $M_2$, 
we can choose a 3-ball $B^3 \subset M_1$ bounded by an $S^2$,
and similarly another 3-ball $B^3 \subset M_2$ bounded by an $S^2$.
We can then paste $M_1$ and $M_2$ along the common $S^2$ 
to obtain another closed 3-manifold $M$, which is denoted by $M_1 \# M_2$
and is known as the connected sum of $M_1$ and $M_2$.
See Fig.~\ref{fig:prime}.
Clearly, $S^3$ is the identity under this operation.

A 3-manifold which cannot be written as a connected sum (except with $S^3$) is known as a \emph{prime} manifold.
Any 3-manifold $M$ is then known to be given uniquely by a connected sum of prime 3-manifolds, \begin{equation}
M=M_1 \# M_2 \# \cdots \# M_k.
\end{equation}
This is known as the prime decomposition of $M$. 

\begin{figure}
\centering
\includegraphics[width=.5\textwidth]{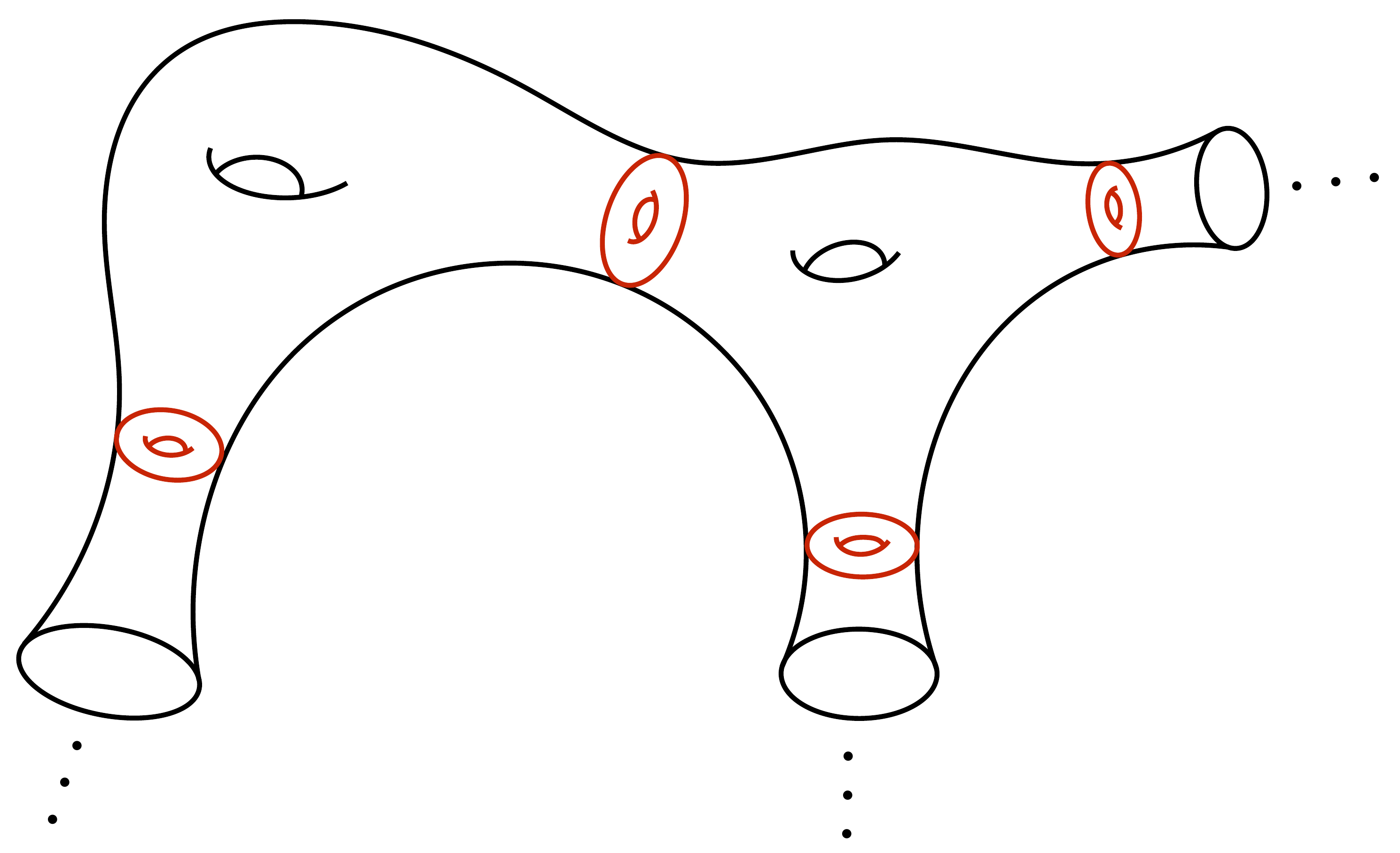}
\caption{Cutting 3-manifolds along $T^2$. \label{fig:torus}}
\end{figure}

\paragraph{Torus decomposition:}

We are now going to cut a prime manifold along embedded $T^2$'s, 
see Fig.~\ref{fig:torus}.
In this case, we keep each piece to have torus boundaries,
since there are multiple distinct ways to fill in a torus boundary by a solid torus.

One can always cut a manifold $M$ along an embedded torus $T^2$ by considering an embedded loop $S^1 \subset M$ and taking its tubular neighborhood. 
If $M$ has a torus boundary, we can also cut it along a $T^2$ slightly inside the boundary.
A manifold $M$ for which the only ways to cut along $T^2$ are either of these two trivial methods is called \emph{atoroidal}.

It is known that we can cut a prime manifold $M$ along copies of $T^2$ so that the resulting pieces are all atoroidal.
This is the torus decomposition of $M$.
In contrast to the prime decomposition, the torus decomposition is known not to be unique,
but the non-uniqueness comes only from well-understood examples.
These results on prime and torus decompositions were established by \cite{JS,J}.

\paragraph{Atoroidal manifolds:}
The next task for us is to understand atoroidal manifolds. 
Their description follows from the geometrization conjecture of Thurston \cite{Thurston},
which states that any 3-manifolds can be cut along $S^2$ and $T^2$ into pieces which have one of eight possible geometries.
In the proof, the Ricci flow of the metric on a given 3-manifold is considered;
physically, we consider the RG flow of the sigma model whose target space is the 3-manifold in question.
Then, the parts with positive curvature shrink, the parts with negative curvature expand. 
Shrinking of the positive curvature parts performs the prime decomposition,
and then the expansion of the negative curvature parts extracts the hyperbolic manifolds.
The remaining connecting parts are shown to be  generically either a $T^2$ fibration
or an $S^1$ fibration,
all of which are known to have one of the eight geometries,
thus proving the conjecture,
whose proof had a long and winding history.
The overall idea of the proof was first indicated in a series of works by Hamilton e.g.~\cite{Hamilton},
and  was later implemented in a series of papers by Perelman starting in \cite{Perelman} and completed by subsequent works.

In the context of the torus decomposition, this means that 
 atoroidal manifolds are either i) Seifert manifolds
or ii) hyperbolic manifolds. 
Here, 
Seifert manifolds are a certain generalization of circle bundles allowing degenerate fibers
which we detail below,
and hyperbolic manifolds are  manifolds of the form $\mathbb{H}^3/\Gamma$,
where $\mathbb{H}^3$ is the three-dimensional hyperbolic space
and $\Gamma$ is a discrete subgroup of its isometry group.
We note that all hyperbolic manifolds are atoroidal, while not all Seifert manifolds are atoroidal.

\begin{figure}
\centering
\includegraphics[width=.5\textwidth]{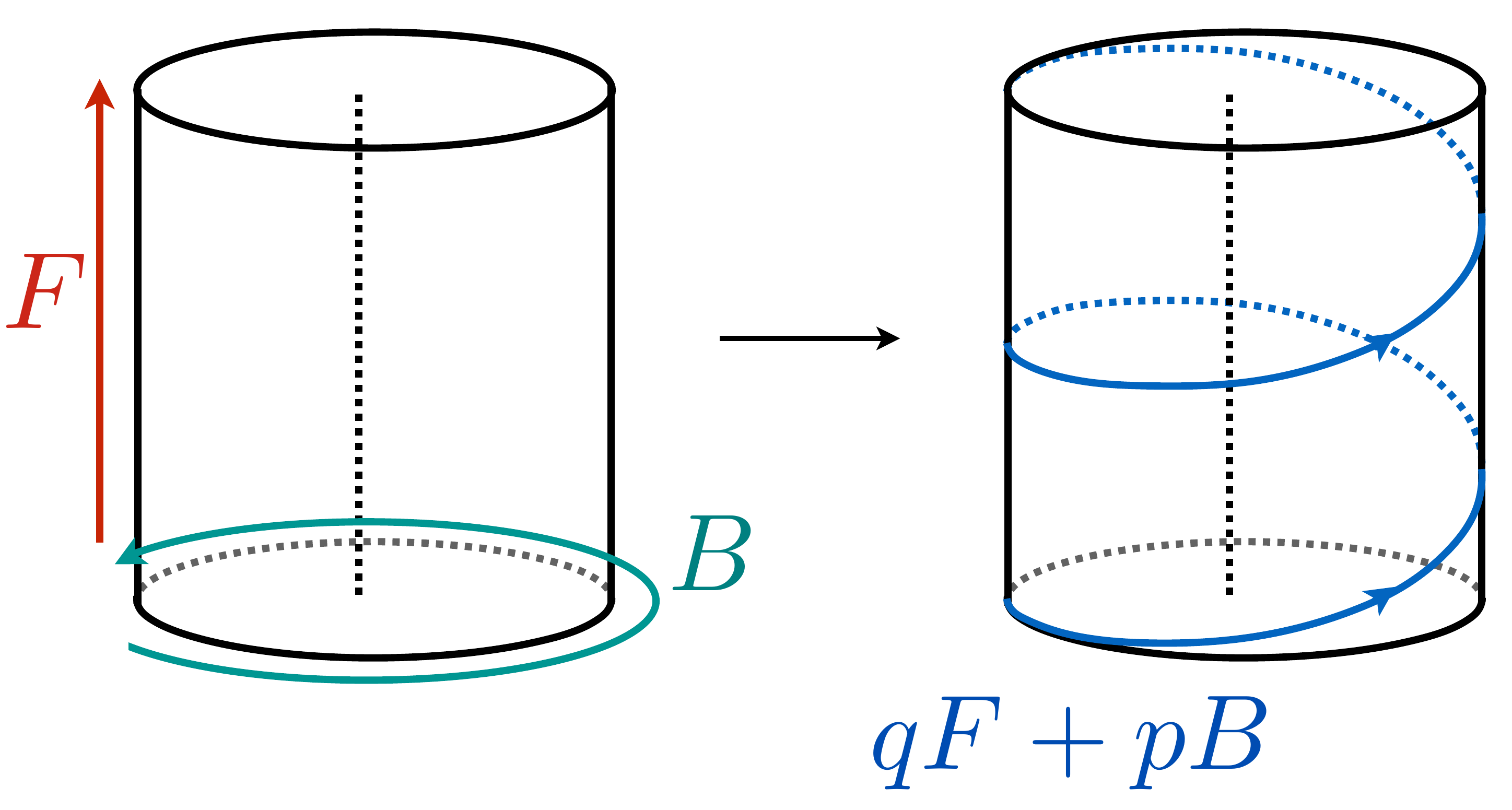}
\caption{Description of a singular fiber of a Seifert fibration. Instead of filling in $B$, one fills in $qF + p B$.  \label{fig:filling}}
\end{figure}

\paragraph{Seifert manifolds:}
In this paper we only consider a subclass of Seifert manifolds.
A 3-manifold $M$ in this subclass has a circle action such that its orbits form a two-dimensional surface $\Sigma$.
The circle action makes $M$ into a circle bundle on generic points of $\Sigma$,
but singular fibers of the following form are allowed.

Namely, starting from an ordinary point in the base $\Sigma$, we consider a neighborhood $D^2$. 
On the solid torus $D^2\times S^1$, we denote $\partial D^2$ by $B$ and the fiber cycle by $F$.
Here the cycle $B$ can be shrunk in the geometry.
We can instead fill in the torus boundary by a solid torus such that 
the 1-cycle $pB+qF$ can be shrunk,
where $p$ and $q$ are coprime.
This operation is known as a Dehn filling.
As part of the definition, we demand that $p$ is nonzero.
This is to guarantee that the circle action is free.
An alternative description consists in gluing back the solid torus $D^2 \times S^1$ by identifying 
the torus boundaries after performing an $SL(2,\bZ)$ action,
whose explicit form will be given in \eqref{explicit}. 

We note that these operations with $p=1$ do not create a singular fiber.
It instead shifts the first Chern class of the circle bundle. 
Therefore, a Seifert manifold of our interest can be constructed
by starting from a trivial $S^1$ bundle over a  base surface of genus $g$, choosing $k$ points on the base, 
and performing the Dehn filling operation above the $i$-th point on the base
with the parameters $q_i/p_i$, producing the $i$-th degenerate fiber.
We will denote this Seifert manifold by the symbol $(g; q_1/p_1,\ldots, q_k/p_k)$;
this is not the standard notation in mathematics literature, but suffices for our purposes.
We also note that it is sometimes assumed that $0\le  q_i < p_i$ to normalize the Seifert parameters.
In this paper we do not assume that, and allow arbitrary pairs $(p_i, q_i)$ as long as $\mathrm{gcd}(p_i,q_i)=1$.


\subsection{Associated 3d theories}
\label{sec:associated}
From the geometric summary above, 
it is clear that we need to understand the following steps 
in order to determine the 3d theories obtained by putting $N$ M5-branes on 3-manifolds.
Namely,
i) we need to understand M5-branes on atoroidal manifolds,
And then, ii) we need to understand what happens 
when we glue two manifolds along $T^2$ or $S^2$.

Let us discuss the question ii) first. 
It should be possible to analyze the gluing along $S^2$,
using the known results concerning the $S^2$ reduction of the 6d \Nequals{(2,0)} theory 
studied e.g.~in \cite{Assel:2016lad}.
As this is not the main focus of this paper,
we will move on to the gluing along $T^2$.
Let us review how it goes.

\subsubsection{Theories on 3-manifolds with torus boundaries}

We need to start by clarifying what is meant by `the 6d \Nequals{(2,0)} theory on 3-manifolds with torus boundaries',
a phrase often used in the context of the 3d/3d correspondence.
Close to a $T^2$ boundary, the manifold can be approximated as $\bR_{>0} \times T^2$.
Compactifying on $T^2$ first, we have 4d \Nequals4 $SU(N)$ super Yang--Mills on $\bR_{>0}$,
where the $SL(2,\bZ)$ transformation acting on $T^2$ is now regarded as the S-duality group action of the \Nequals4 super Yang--Mills.

We can give a boundary condition at the boundary of $\bR_{>0}$ 
by choosing a duality frame for the \Nequals4 super Yang--Mills,
and taking the Dirichlet condition in the chosen frame.
This gives rise to an $SU(N)$ flavor symmetry.
To choose a duality frame, we need to specify an electric 1-cycle $E$ and a magnetic 1-cycle $M$ on $T^2$, forming a basis of $H_1(T^2,\bZ)$, such that 
self-dual strings wrapped around $E$ correspond to the W-bosons for the $SU(N)$ flavor symmetry at the boundary.
Summarizing, to associate a 3d theory to a 3-manifold with torus boundaries,
we need to fix a basis of $H_1(T^2,\bZ)$ at each boundary.

The necessity of this additional data at the boundary can be understood also as follows.
The 6d \Nequals{(2,0)} theory is a chiral theory
and cannot be placed on a space with boundary.
To make a closed manifold, we can paste to a torus boundary the product manifold
\begin{equation}
S^1\times \text{($D^2$ with a full puncture at the origin)}.
\end{equation}
Then $S^1$ is the magnetic 1-cycle $M$ and $\partial D^2$ is the electric 1-cycle $E$
in the description above.

We still find it convenient to assign a 3d theory to 3-manifolds with torus boundaries, 
since such open manifolds commonly appear in the description of classification of 3-manifolds.
When we refer to such a 3d theory, we always make a choice of the magnetic 1-cycle $M$
and the electric 1-cycle $E$ at the torus boundary,
and implicitly insert the full puncture along the magnetic 1-cycle $M$, 
 shrinking the electric 1-cycle $E$ in the process.

\subsubsection{Gluing along $T^2$}

Given two $T^2$ boundaries with the same choice of 1-cycles $E$ and $M$,
we can simply glue them by gauging the diagonal combination of 
two $SU(N)$ symmetries associated to the two $T^2$ boundaries
by 3d \Nequals4 $SU(N)$ vector multiplet. 
In order to glue two $T^2$ boundaries with two different choices of pairs of 1-cycles $E$ and $M$,
we need to perform the $SL(2,\bZ)$ transformation first. 
Any $SL(2,\bZ)$ transformation can be done by a sequence of $S$ and $T$ transformations given by \begin{equation}
S=\begin{pmatrix}
0 & -1 \\
1 & 0
\end{pmatrix},\qquad
T=\begin{pmatrix}
1 & 1 \\
0 & 1
\end{pmatrix},
\end{equation}
whose manifestation as operations on 3d supersymmetric theories was determined in \cite{Gaiotto:2008sd,Gaiotto:2008ak}.
Namely, the $S$ transformation is realized by coupling to the duality wall theory $T(SU(N))$,
and the $T$ transformation is realized by shifting the Chern--Simons level of the $SU(N)$ flavor background by one.

The $S$ transformation preserves 3d \Nequals4 supersymmetry, 
while the $T$ transformation preserves only 3d \Nequals3 supersymmetry,
since it involves the supersymmetric Chern--Simons coupling.
Therefore, gluing with an arbitrary $SL(2,\bZ)$ transformation always preserves at least 3d  \Nequals3 supersymmetry.

\subsubsection{Ingredients} 
\label{sec:ingredients}
Now that we have some understanding of the gluing operations,
we need to understand the ingredients we glue, i.e.~theories associated to atoroidal manifolds.
As reviewed above, atoroidal manifolds are either hyperbolic manifolds or Seifert manifolds,
which we discuss in turn.

\paragraph{Theories on hyperbolic manifolds:}
The 3d theories obtained by putting \emph{two} M5-branes on hyperbolic manifolds were the focus of the epoch-making paper \cite{Dimofte:2011ju,Dimofte:2011py}.
The theories constructed in these original papers, however, associates a $U(1)$ flavor symmetry for each $T^2$ boundary, rather than an $SU(2)$ flavor symmetry, 
which we expect on general grounds.
Such theories with an $SU(2)$ flavor symmetry per $T^2$ boundary were later studied in \cite{Chung:2014qpa,Gang:2018wek}.
The generalization to more than two M5-branes has also been studied, starting in \cite{Dimofte:2013iv}.

\paragraph{Theories on Seifert manifolds:} 
The 3d theories obtained by putting $N$ M5-branes on Seifert manifolds have also been determined previously, see e.g.~\cite{Gadde:2013sca,Chung:2014qpa,Pei:2015jsa,Gukov:2016gkn,Gukov:2017kmk,Eckhard:2019jgg,Cho:2020ljj},
although the determination was usually a byproduct rather than the main topic of these papers.
Here we give a brief summary of the construction.

As recalled above, Seifert manifolds of our interest are obtained by starting from a direct product of $S^1$ and a surface $\Sigma$ with boundaries,
and performing a Dehn filling for each of its boundaries.
A Dehn filling is done by gluing an $S^1\times D^2$ after an appropriate $SL(2,\bZ)$ transformation.
As $D^2$ is also a special case of a surface with boundaries, 
all that is left is to understand the 3d theory obtained by putting $N$ M5-branes on a product 
$S^1\times\Sigma$, with $\Sigma$ a surface with $m$ boundaries $\partial\Sigma = C_1\cup\cdots\cup C_m$.

\paragraph{Theories on $S^1\times \Sigma$: }
Let us first pick $C_i$ to be the electric cycle and the $S^1$ fiber to be the magnetic cycle.
Equivalently, let us  fill in $C_i$ by a disk $D^2_i$ with a full puncture at the origin.
We then have $N$ M5-branes on a product $S^1$ times a sphere $S^2$ with $m$ full punctures.
This then gives the $T_{N,m}$ theory we already used in Sec.~\ref{sec:mainexamples}.

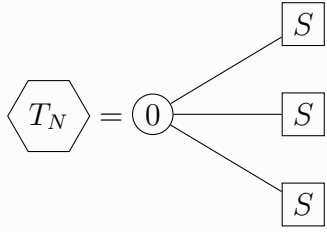
\begin{figure}
\[
\begin{tikzpicture}[baseline=(TN.base)]
\node[hexagon] (TN) at (0,0) {$T_N$};
\end{tikzpicture}
=
\begin{tikzpicture}[baseline=(Z.base)]
\node[circ,inner sep=2pt] (Z) at (0,0) {$0$};
\node[draw,rectangle] (A) at (2,1.2) {$S$};
\node[draw,rectangle] (B) at (2,0) {$S$};
\node[draw,rectangle] (C) at (2,-1.2) {$S$};
\draw (Z)--(A);
\draw (Z)--(B);
\draw (Z)--(C);
\end{tikzpicture}
\]
\caption{The 3d mirror of the $T_N$ theory is obtained by taking three copies of $T(SU(N))$ theory and gauge its diagonal $SU(N)$ symmetry.
The circle with zero in it symbolizes the gauging with zero Chern--Simons level, 
i.e.~the \Nequals4 gauging.\label{fig:TN}}
\end{figure}

Secondly, we can choose $C_i$ to be the magnetic cycle and the $S^1$ fiber to be the electric cycle.
Equivalently, let us fill in $S^1$ of each boundary $S^1\times C_i$ by a $D^2$, again with a full puncture at the origin.
The two ways of filling are related by the S-transformation at each boundary,
and it is known that the second method leads to the 3d `theory' which is simply the diagonal Dirichlet boundary condition
setting all $SU(N)$ flavor symmetries associated to $m$ boundaries the same \cite[Sec.~5.2]{Benini:2010uu}.\footnote{%
This is analogous to the situation that happens when the two $T(SU(N))$ theories are coupled via a diagonal gauging.
The resulting 3d `theory' formally has two $SU(N)$ symmetries, but in fact acts as a `delta function' forcing the two sets of background fields coupled to two $SU(N)$ symmetries to be the same, 
i.e.~it is the diagonal Dirichlet boundary condition.
In fact this is exactly the case $m=2$ in the discussion in the main text.
}
These two descriptions are related by the 3d mirror symmetry.
Summarizing, the $T_{N,m}$ theory is given by taking $m$ copies of $T(SU(N))$ theories with $SU(N)^{(i)}_H\times SU(N)_C^{(i)}$ flavor symmetry, for $i=1,\ldots,m$,
and gauging the diagonal subgroup $SU(N)_C$ of $SU(N)_C^{(1,\ldots,m)}$.
We illustrate the case of $T_N=T_{N,3}$ in Fig.~\ref{fig:TN}.
The genus-$g$ version is known to be obtained simply by adding $g$ adjoint hypermultiplets 
when we gauge $SU(N)_C$.
We denote the resulting theory by $T_{N,m,g}$.

We can use this description of the $T_{N,m,g}$ theory to derive our crucial chiral ring relation \eqref{superimportant}, using some basic properties of the $T(SU(N))$ theory.
Denote the moment map fields of its $SU(N)_H$ and $SU(N)_C$ flavor symmetries
by $\mu_H$ and $\mu_C$ respectively.
Let us couple $T(SU(N))$ with adjoint scalar fields $\Phi_C$, $\Phi_H$ via  \begin{equation}
W= \tr \Phi_H \mu_C + \tr \Phi_C \mu_H.
\end{equation}
Then  the following crucial chiral ring relations:
\begin{equation}
\tr (\Phi_H)^n = \tr (\mu_H)^n, \qquad
\tr (\Phi_C)^n = \tr (\mu_C)^n \label{Phimu}
\end{equation} 
are satisfied for arbitrary $n$.

One way to derive these relations is to realize the $T(SU(N))$ theory as 4d \Nequals4 super-Yang--Mills on a segment with suitable boundary conditions, as in \cite{Gaiotto:2008ak}.
The undeformed $T(SU(N))$ theory is known to be the S-dual of the Nahm pole boundary condition \begin{equation}
\phi(x) \sim \frac{\nu}x
\end{equation} where $x$ is the distance to the boundary, 
 $\nu$ is the principal nilpotent element in $\mathfrak{su}(N)$,
 and $\sim$ denotes that the two are conjugate.
As $\phi(x)$ at the other boundary $x=x_0$ is identified with $\mu_H$, 
we see that $\tr (\mu_H)^n=0$ for all $n$.

A nonzero diagonalizable $\Phi_H$ is known to deform this Nahm pole boundary condition 
to the form \begin{equation}
\phi(x) \sim \Phi_H + \frac{\nu'} x,
\end{equation} 
where $\nu'$ is now the principal nilpotent element in the commutant of $\Phi_H$ in $\mathfrak{su}(N)$.
From this we see $\tr (\Phi_H)^n = \tr (\mu_H)^n$ for all $n$.
The equation $\tr (\Phi_C)^n = \tr (\mu_C)^n$ follows perfectly analogously.

With this property of the $T(SU(N))$ theory in hand, it is a simple matter to derive the chiral ring relation \eqref{superimportant}.
The $T_{N,m}$ theory is given by taking 
$m$ copies of the $T(SU(N))$ theory and coupling them to a single \Nequals4 $SU(N)$ vector multiplet,
with the superpotential \begin{equation}
W=\sum_i \tr \Phi \mu_C^{(i)},
\end{equation}
where $\Phi$ is now dynamical. This immediately implies \begin{equation}
\tr (\mu_H^{(i)})^k = \tr \Phi^k,
\end{equation}  independent of $i$ for all $k$.
As $\mu_H^{(i)}$ are identified as the moment map field $\mu_i$ of the $i$-th $SU(N)$ of the $T_{N,m}$ theory, 
we derived \eqref{superimportant}, that is, \begin{equation}
\text{$\tr (\mu_i)^2$ is independent of $i$}.
\end{equation}

\paragraph{Dehn fillings:}
To complete our description of the basic operations, let us briefly discuss
how to perform the Dehn fillings of a torus boundary.
Geometrically, it is done by pasting $S^1\times D^2$ along the torus boundary.
We can associate multiple 3d theories to $S^1\times D^2$ related by $SL(2,\bZ)$ operations
depending on the choice of a basis of $H_1(T^2,\bZ)$ at its boundary.
Equivalently, these are obtained by pasting $S^1\times D^2$ and $S^1\times (\text{$D^2$ with a full puncture})$ with $SL(2,\bZ)$ operations along their torus boundaries.

These are the special $m=1$ case of what we just discussed above.
In particular, when the $SL(2,\bZ)$ operation is trivial, the 3d manifold is $S^1\times S^2$ with a full puncture wrapping $S^1$, and the associated theory is $T(SU(N))$ with one of its  $SU(N)$ symmetry gauged. This is the $T_{N,1}$ theory.
In contrast, when we use the $S$ operation, the 3d manifold is $S^3$,
with a full puncture wrapping a circle fiber of its Hopf fibration. 
This is a rather degenerate setup, and simply gives a free boundary condition to the background fields of the $SU(N)$ flavor symmetry supported on the puncture. 

In the language of the electric/magnetic 1-cycle chosen at the torus boundary,
this means that filling in the magnetic cycle is simply done by gauging the $SU(N)$ flavor symmetry with an \Nequals4 vector multiplet,
and that filling in the electric cycle is done by coupling it to the $T_{N,1}$ theory,
again via an \Nequals4 $SU(N)$ vector multiplet.

\subsection{Main examples from geometry}
\label{sec:moreexamples}
As an example, let us consider the 3d theory associated to a Seifert bundle over $S^2$ 
with three singular fibers, with Seifert parameters $1/k_{1,2,3}$, respectively.
This manifold is obtained by first considering $S^1$ times $S^2$ with three holes whose boundaries we denote by $C_{1,2,3}$,
and Dehn-filling three torus boundaries by filling the direction $F+k_i C_i$ 
where $F$ is the fiber $S^1$ direction. 

As discussed, the 3d $T_N$ theory corresponds to the choice of $F$ as the magnetic 1-cycle and $C_i$ as the electric 1-cycle.
For the solid torus $S^1\times D^2$, 
we define (perhaps surprisingly) $C'=S^1$ and $F'=\partial D^2$.
We then specify the gluing in terms of the $SL(2,\bZ)$ matrix $g$ appearing in 
the relation \begin{equation}
\label{eq:peculiar}
(C, F) g
= (C' , F') .
\end{equation}
This somewhat peculiar convention of ours is designed so that the Dehn filling with the identity element $1\in SL(2,\bZ)$ is done by a simple gauging of the $SU(N)$ flavor symmetry.\footnote{%
Note that in 4d \Nequals2 class S constructions, 
a simple gauging of the $SU(N)$ flavor symmetry associated to a full puncture
without coupling to another sector 
corresponds to converting the full puncture to an irregular puncture of an appropriate type.
Its $S^1$ compactification does not directly equal the Dehn filling with $g=1$ discussed here.
}

In our case we  have $g =T^{k_i}$,
which simply adds the level $k_i$ Chern--Simons coupling.
This means that the resulting 3d theory
is  the $T_N$ theory whose three $SU(N)$ flavor symmetries
are gauged with 3d \Nequals3 Chern--Simons coupling $k_i$.

Another description is as follows. 
We need to perform Dehn fillings at the three torus boundaries.
Our choice of the Seifert parameters means that at the $i$-th boundary,
we perform the $SL(2,\bZ)$ operation $T^{k_i}S$,
before gluing in the $T_{N,1}$ theory. 
As the $T_{N,1}$ theory is itself the $T(SU(N))$ theory with $SU(N)_C$ gauged,
two $S$ operations cancel,
and we end up simply gauging the $i$-th $SU(N)$ flavor symmetry of the $T_N$ theory
with the \Nequals3 Chern--Simons term with the level $k_i$.

More generally, when the Seifert parameters are $q_i/p_i$,
one needs to perform an appropriate $SL(2,\bZ)$ transformation to bring 
the cycle $q_i F+p_i C_i$ 
to be shrunk into the magnetic 1-cycle. 
This can be done by the transformation 
\begin{equation}
g=T^{k^{(1)}} S T^{k^{(2)}} S \cdots S T^{k^{(a-1)}} ST^{k^{(a)}} \label{explicit}
\end{equation}
where the Chern--Simons levels $k^{(t)}$ for $t=1,2,\ldots,a$ is given by 
a continued fraction expansion of the Seifert parameters, i.e.~by \begin{equation}
\frac{q}{p}=\dfrac{1}{k^{(1)} - \dfrac{1}{k^{(2)}-\dfrac{1}{k^{(3)}-\cdots}}}.
\end{equation}
Note that we dropped the subscript $i$ to specify the singular fiber considered, to lighten the notation somewhat.

\def\LINE#1#2#3#4{
\node[circ] (k1) at (2,#1) {$k_{#3}^{(1)}$};
\node[draw,rectangle] (n1) at (3,#1) {$S$};
\node[circ] (k1x) at (4,#1) {$k_{#3}^{(2)}$};
\node[draw,rectangle] (n1x) at (5,#1) {$S$};
\node[draw,rectangle] (n1y) at (7,#1) {$S$};
\node[circ] (#2) at (8,#1) {$k_{#3}^{(a_{#3})}$};
\draw (#4)--(k1)--(n1)--(k1x)--(n1x);
\draw[dashed] (n1x)--(n1y);
\draw (n1y)--(#2);
}
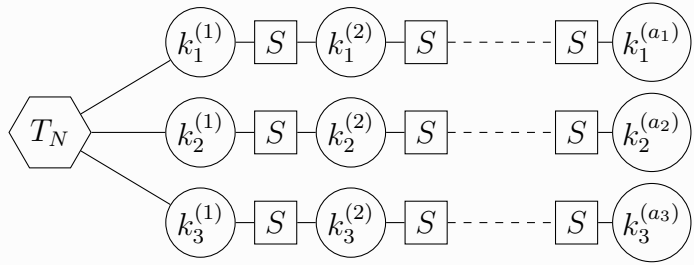
\begin{figure}
\centering
\begin{tikzpicture}
\node[hexagon] (TN) at (0,0) {$T_N$};
\LINE{1.2}{A}1{TN}
\LINE{0}{B}2{TN}
\LINE{-1.2}{C}3{TN}
\end{tikzpicture}
\caption{The 3d theory associated to the Seifert bundle over $S^2$ with three singular fibers.
For the notation, see the main text. \label{fig:theory}}
\end{figure}

Summarizing, we see that the 3d theory obtained from the Seifert bundle over $S^2$ with three punctures with parameters $q_i/p_i$ has the form given in Fig.~\ref{fig:theory}.
There, a circle enclosing $k$ denotes an \Nequals3 Chern--Simons  vector multiplet with level $k$,
a hexagon is the 3d $T_N$ theory,
and a box with an $S$ denotes the 3d $T(SU(N))$ theory.
This is a well-known result \cite{Gadde:2013sca,Pei:2015jsa,Gukov:2016gkn,Gukov:2017kmk,Eckhard:2019jgg,Cho:2020ljj},
although the generic presence of \Nequals3 supersymmetry was not explicitly remarked in those papers.

In the simplest case when $q_i/p_i=1/k_i$,
these are exactly the theories considered in Sec.~\ref{sec:mainexamples},
where we saw that the supersymmetry enhances to \Nequals4 when $1/k_1+1/k_2+1/k_3=0$.
Our question now is whether we can understand this enhancement geometrically.
To study this question, we need to come back to a general question on
how the geometry determines the amount of supersymmetries preserved.

\subsection{Number of supersymmetries from the geometry}
\label{sub:number-seifert}

The 6d theory on M5-branes has the  R-symmetry $Sp(2)_R\simeq SO(5)_R$.
A generic 3-manifold $M$ has holonomies in $SO(3)$.
We can turn the R-symmetry background on $M$, by embedding the $SO(3)$ spin connection into the R-symmetry by 
\begin{equation}
SO(3)\subset SO(3) \times SO(2) \subset SO(5)_R.\label{so3embedding}
\end{equation}
A short computation reveals that two supercharges remain,
realizing 3d \Nequals2 supersymmetry.\footnote{There is another way to to embed $SO(3)$ into $SO(5)$, namely 
$SO(3) \subset SO(3) \times SO(3) \subset SO(4) \subset SO(5)$ (at the level of the Lie algebra). 
This alternative twist preserves only 3d \Nequals{1} supersymmetry.}

As another example, let us suppose the 3-manifold $M$ is a product $S^1\times \Sigma$.
As is well-known, M5-branes on $\Sigma$ give a 4d \Nequals2 theory,
and a further compactification on $S^1$ gives a 3d \Nequals4 theory.
More generally, any 3-manifold $M$ whose holonomies are in $SO(2)$ gives rise to a 3d \Nequals4 theory.
Indeed, we can turn the R-symmetry background on $M$ by embedding the $SO(2)$ spin connection to the R-symmetry by 
\begin{equation}
SO(2)\subset SO(5)_R.
\end{equation}
Again a short computation reveals that four supercharges remain,
realizing 3d \Nequals4 supersymmetry. 

Therefore, a way to look for the cause of the \Nequals4 enhancements 
is to study when a 3-manifold has a reduced holonomy in $SO(2)$.
A simple result which is crucial to us is the following statement:
\begin{claim}
\textbf{Theorem:} A Seifert manifold of type $(g; q_1/p_1,\ldots, q_m/p_m)$ in our notation 
is, unless $g=0$ and $m<3$,
 an order-$n$ quotient of an $S^1$ fibration over a Riemann surface $\Sigma$,
where $n=\mathrm{lcm}\, \{p_i\}$
and the first Chern class of the $S^1$ fibration is $n\sum q_i/p_i$.
The quotient is given by an order-$n$ isometry of $\Sigma$ 
combined with a $1/n$ rotation of the fiber.
\end{claim}
This theorem follows from e.g.~Lemma 3.7 of \cite{Scott},
where the same statement is proved except that the quotient can be taken to be an isometry.
That the quotient can be taken to be an isometry follows by first considering an arbitrary metric
and then averaging it over the quotient group.

Now, this theorem in particular implies that a Seifert manifold of type $(g; q_1/p_1,\ldots, q_k/p_k)$
is an order-$n$ quotient of a trivial product $\Sigma\times S^1$
when $\sum_i q_i /p_i =0$.
As the quotient process in this case does not affect the holonomy group $SO(2)$,
we see that the 3d theories associated to this class of Seifert manifolds
should have an enhanced \Nequals4 supersymmetry
when $\sum_i q_i/p_i=0$.

This gives a geometric explanation of the \Nequals4 supersymmetry 
of the theories we introduced in Sec.~\ref{sec:mainexamples},
namely the $T_N$ theory gauged with $SU(N)^3$ with levels $k_{1,2,3}$ 
when $1/k_1+1/k_2+1/k_3=0$.
This is because, as we saw in the subsection above,
they come from Seifert manifolds of type $(0;1/k_1,1/k_2,1/k_3)$.
This analysis can be easily generalized to the $T_{N,m,g}$ theory
whose $SU(N)^m$ symmetry is gauged with Chern--Simons levels $k_i$.
It enhances to \Nequals4 
when $\sum 1/k_i=0$.
This agrees with the geometric theorem quoted above,
since the corresponding Seifert manifold is of type $(g;1/k_i)$.

Our geometric analysis also predicts that the 3d theories given in Sec.~\ref{sec:moreexamples}
should have an enhanced \Nequals4 supersymmetry when $\sum q_i/p_i=0$.
This is the task we would like to investigate next.
Before that, we need to establish a field-theoretical result first.


\subsection{A field-theoretical interlude}
\label{sec:interlude}

\subsubsection{The statement to be established}
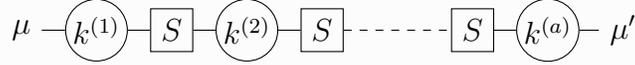
\begin{figure}
\centering
\begin{tikzpicture}
\node (m1) at (1,1.2) {$\mu$};
\node[circ] (k1) at (2,1.2) {$k^{(1)}$};
\node[draw,rectangle] (n1) at (3,1.2) {$S$};
\node[circ] (k1x) at (4,1.2) {$k^{(2)}$};
\node[draw,rectangle] (n1x) at (5,1.2) {$S$};
\node[draw,rectangle] (n1y) at (7,1.2) {$S$};
\node[circ] (k1y) at (8,1.2) {$k^{(a)}$};
\node (mm) at (9,1.2) {$\mu'$};
\draw (m1)--(k1)--(n1)--(k1x)--(n1x);
\draw[dashed] (n1x)--(n1y);
\draw (n1y)--(k1y)--(mm);
\end{tikzpicture}
\caption{A single chain of $T(SU(N))$ theories coupled by \Nequals3 Chern--Simons multiplets. \label{fig:chain}}
\end{figure}

To analyze the theory depicted in Fig.~\ref{fig:theory},
the part essential to us is the chain structure shown in Fig.~\ref{fig:chain} implementing 
an $SL(2,\bZ)$ transformation $g$.
There,  
the lines connecting to $\mu$ and $\mu'$ on the extreme left and right
signify that the first and the last \Nequals3 Chern--Simons multiplets
also couple to these complex adjoint scalar fields,
and the Chern--Simons levels $k^{(i)}$ are determined in terms of the continued fraction expansion \begin{equation}
\frac{q}{p}=\dfrac{1}{k^{(1)} - \dfrac{1}{k^{(2)}-\dfrac{1}{k^{(3)}-\cdots}}}\label{cfrac}
\end{equation}
as before.
We note that $g$ has the form \begin{equation}
g= T^{k^{(1)}} S T^{k^{(2)}} S \cdots S T^{k^{(a)}}=\begin{pmatrix}
q' & p \\
r & q
\end{pmatrix}\label{g}
\end{equation} where \begin{equation}
\frac{q'}{p}=\dfrac{1}{k^{(a)} - \dfrac{1}{k^{(a-1)}-\dfrac{1}{k^{(a-2)}-\cdots}}}.
\end{equation}

Let us denote by $\mu_{C,H}^{(i)}$ the Coulomb-branch and Higgs-branch moment map fields of the $i$-th $T(SU(N))$ theory,
and by $\Phi^{(i)}$  the adjoint scalar of the $i$-th Chern--Simons multiplet.
The superpotential is then given by \begin{equation}
W= \sum_{i=1}^{a}\left(\tr \mu_H^{(i-1)} \Phi^{(i)} -\frac{k^{(i)}}2\tr (\Phi^{(i)})^2 + \tr\Phi^{(i)} \mu_C^{(i)}\right)   
\end{equation}
where  $\mu_H^{(0)}:=\mu$ and $\mu_C^{(a)} := \mu'$.

We now show the following statement inductively in terms of the length $a$ of the chain: 
\begin{claim}
\textbf{Statement:} The effective superpotential is equivalent to \begin{equation}
W = \frac12 \frac{q}{p} \tr \mu^2  
+ \sum_{i=1}^{a}c^{(i)}\tr \mu_H^{(i-1)}\mu_C^{(i)} 
+\frac12\frac{q'}{p} \tr ( \mu' )^2,
\label{W}
\end{equation}
after eliminating $\Phi^{(j)}$. 
Here,  $c^{(i)}$ are certain constants satisfying \begin{equation}
\prod c^{(i)} = 1/p.
\end{equation}
\end{claim}
Those readers who are not interested in the (rather technical) derivation can skip to the beginning of the next section \ref{sec:enh}.

\subsubsection{A lemma and a corollary}

Before we start the derivation of the statement above, we need the following lemma:
\begin{claim}
\textbf{Lemma:} For a single $T(SU(N))$ theory, the deformation \begin{equation}
W=\tr \Phi \mu_C + \tr \mu_H \Phi' + \frac{a}2 \tr(\mu_H)^2, \label{1st-deformation}
\end{equation}
and the deformation \begin{equation}
W= \frac{a}2\tr \Phi^2 + \tr \Phi \mu_C + \tr \mu_H\Phi'\label{2nd-deformation}
\end{equation}
are equivalent at finite nonzero $a$,
where $\Phi$ and $\Phi'$ are  background superfields.
\end{claim}
To show this, we consider how the  chiral ring relation at $a=0$,\begin{equation}
\tr\Phi^2=\tr(\mu_H)^2 \label{cr}
\end{equation}
which we already recalled at \eqref{Phimu},
would be deformed under the first deformation \eqref{1st-deformation} at nonzero $a$.
We assign \Nequals2 R-charge $+1$ to $\Phi$, $\Phi'$, $\mu_{C,H}$;
we also assign the $J_{34}$ charge $+1$ to $\Phi$, $\mu_H$ and $-1$ to $\Phi'$, $\mu_C$.
Then $a$ has the R-charge $0$ and the $J_{34}$-charge $-2$.
As the equation has the R-charge $2$ and the $J_{34}$-charge $2$,
the only term one can write is $a^{-2}$ times a quadratic combination of $\Phi'$ and $\mu_C$.
This is forbidden, since the chiral ring relation should be continuous at $a=0$.
This establishes that the relation \eqref{cr} holds unchanged under the deformation \eqref{1st-deformation} at finite $a$. 
This means that 
the deformation \eqref{1st-deformation} is equivalent to \begin{equation}
W=\frac\epsilon 2\tr\Phi^2 + \tr \Phi \mu_C + \tr \mu_H \Phi' + \frac{1}{2}(a-\epsilon) \tr(\mu_H)^2
\end{equation} for an infinitesimal $\epsilon$.
This implies that it is further equivalent  to \eqref{2nd-deformation} and concludes the derivation of the lemma.

Let us now add a term $\tr \tilde \Phi \Phi - \frac k2 \tr \Phi^2 $ to both \eqref{1st-deformation} and \eqref{2nd-deformation},  integrate out $\Phi$,
and replace the resulting $\tilde \Phi$ by $\Phi$ since the original $\Phi$ no longer appears.
We immediately obtain the following corollary: 
\begin{claim}
\textbf{Corollary:} 
For a single $T(SU(N))$ theory, the deformation \begin{equation}
W'=\frac {1}{2k} \tr ( \Phi + \mu_C)^2 + \frac a2 \tr(\mu_H)^2 + \tr \mu_H \Phi' 
\end{equation} and 
\begin{equation}
W'= \frac {1}{2(k-a)} \tr ( \Phi + \mu_C)^2  + \tr \mu_H\Phi'
\end{equation}
are equivalent at finite nonzero $a$ and $k$,
where $\Phi$ and $\Phi'$ are  background superfields.
\end{claim}

We note that  the lemma and the corollary only guarantee
that the two deformations 
are equivalent up to $\bar D_{\dot\alpha}$-exact terms.
In the case we encounter below, $\Phi$ and $\Phi'$ are either chiral primaries
from SCFTs other than the $T(SU(N))$ theory in question. 
Therefore, the $\bar D_{\dot\alpha}$-exact terms are actually absent.
The same can also be said when we actually use the corollary.

\subsubsection{Derivation}

Let us begin the derivation of the statement itself. 
The case  when the length is $1$ is immediate.
Let us then assume that this statement has been shown up to the length $a-1$.
We see that the superpotential is equivalent to \begin{equation}
W = \frac{1}{2k^{(1)} }\tr (\mu + \mu_C^{(1)})^2 
+\frac12\frac{\tilde q}{\tilde p} \tr (\mu_H^{(1)})^2 
+\sum_{i=2}^{a}c^{(i)}\tr \mu_H^{(i-1)}\mu_C^{(i)} 
+ \frac12\frac{\tilde q'}{\tilde p} \tr (\mu')^2 
\end{equation}
where \begin{equation}
\tilde g=  T^{k^{(2)}} S \cdots S T^{k^{(a)}}=\begin{pmatrix}
\tilde q' & \tilde p \\
\tilde r & \tilde q
\end{pmatrix}.
\end{equation}

Applying the corollary above, we can rewrite $W$ as   \begin{equation}
W=  \frac12\frac{q}{p} \tr (\mu+\mu_C^{(1)})^2
+\sum_{i=2}^{a}c^{(i)}\tr \mu_H^{(i-1)}\mu_C^{(i)} 
+ \frac12\frac{\tilde q'}{\tilde p} \tr (\mu')^2
\label{ABABAB}
\end{equation}
where we used $p/q=k^{(1)}-\tilde q/\tilde p$.
We now repeatedly use the lemma applied to the $i$-th $T(SU(N))$ theory,
and perform the following replacement of the superpotential deformations:
\begin{equation}
\tr(\mu_C^{(i)})^2 \leftrightarrow (c^{(i+1)})^2 \tr(\mu_C^{(i+1)})^2.
\end{equation} 
More explicitly, we do the exchanges \begin{multline}
\frac12\frac qp \tr (\mu_C^{(1)})^2
\leftrightarrow \frac12\frac qp (c^{(2)})^2 \tr (\mu_C^{(2)})^2
\leftrightarrow \frac12\frac qp (c^{(2)}c^{(3)})^2 \tr (\mu_C^{(3)})^2 \\
\leftrightarrow\cdots
\leftrightarrow \frac12 \frac qp \left(\prod_{i=2}^a c^{(a)}\right)^2 \tr (\mu')^2
=
\frac12\frac{q}{p}\frac{1}{\tilde p^2} \tr(\mu')^2
\end{multline} where we used the inductive assumption in the last equality.
This allows us to rewrite \eqref{ABABAB}  further to \begin{equation}
W=  \frac12\frac{q}{p} \tr \mu^2  +
\frac qp \tr \mu\mu_C^{(1)}  
+ \sum_{i=2}^{a}c^{(i)}\tr \mu_H^{(i-1)}\mu_C^{(i)} 
+ \frac12\left(\frac{q}{p}\frac{1}{\tilde p^2}+\frac{\tilde q'}{\tilde p}\right) \tr (\mu')^2.
\end{equation}
As \begin{equation}
\begin{pmatrix}
q' & p \\
r & q
\end{pmatrix}
= T^{k^{(1)}} S \begin{pmatrix}
\tilde q' & \tilde p\\
\tilde r & \tilde q
\end{pmatrix},
\end{equation} we have \begin{equation}
q=\tilde p, \qquad r=\tilde q'.
\end{equation}
Using these relations, we finally find that \begin{equation}
W = \frac12 \frac{q}{p} \tr \mu^2  
+ \sum_{i=1}^{a}c^{(i)}\tr \mu_H^{(i-1)}\mu_C^{(i)} 
+\frac12\frac{q'}{p} \tr ( \mu' )^2,
\end{equation}
with \begin{equation}
\prod_{i=1}^a c^{(i)} = 1/p,
\end{equation}
which is exactly the property we wanted to demonstrate for the length $a$ chain.\footnote{%
Note that the constants $c^{(i)}$ apparently depends on the order of the integrating-out of $\Phi^{(i)}$. 
For example, when we integrate out $\Phi^{(a)}$ first and $\Phi^{(1)}$ last,
$c^{(i)}=1/(k^{(i)} - 1/(k^{(i+1)}-\cdots))$,
whereas when we integrate out $\Phi^{(1)}$ first and $\Phi^{(a)}$ last, we obtain
$\tilde c^{(i)} = 1/(k^{(i)} - 1/(k^{(i-1)}-\cdots))$ instead.
We still have $\prod c^{(i)} =\prod \tilde c^{(i)} = 1/p$,
and indeed that is the only invariant information in the effective superpotential $W$.
This is because the $i$-th $T(SU(N))$ theory has the symmetry $J_{34}^{(i)}$, under which
$\mu_C^{(i)}$ has charge $-2$ and 
$\mu_H^{(i)}$ has charge $+2$.
This allows us to rescale $(c^{(i)},c^{(i+1)}) \mapsto ( c^{(i)}/s^{(i)} ,s^{(i)} c^{(i+1)})$
for each $i$, under which the only invariant combination is $\prod c^{(i)}$.
YT thanks discussions over Twitter on this point, notably 
\url{https://twitter.com/Coo_Butsukou/status/1549638950897483776},
\url{https://twitter.com/END_OF_PAIOTU/status/1549655415860101125},
\url{https://twitter.com/mathraphsody/status/1549655393735176194}.
}
We make a further remark on the $SL(2,\bZ)$ duality wall theory in Appendix~\ref{app:wall},
using the Statement established above.

\subsection{Supersymmetry enhancements accounted for by holonomy}
\label{sec:enh}

We now analyze the structure of the theory shown in Fig.~\ref{fig:theory}.
Namely, we take three tails of the form discussed in \eqref{W}, and couple them to a $T_N$ theory. 
After eliminating the adjoint scalars in the \Nequals3 Chern--Simons multiplets, 
we have the superpotential of the form \begin{equation}
\begin{aligned}
W= & \frac12 \frac{q_1}{p_1} \tr (\mu_1)^2 + \frac{q_1}{p_1}  \tr \mu_1 \mu_{1,C}^{(1)} + \sum_{i=2}^{a_1-1}c_1^{(i)}\tr \mu_{1,H}^{(i-1)}\mu_{1,C}^{(i)} \\
&+\frac12 \frac{q_2}{p_2} \tr (\mu_2)^2 + \frac{q_2}{p_2}  \tr \mu_2 \mu_{2,C}^{(1)} + \sum_{i=2}^{a_2-1}c_2^{(i)}\tr \mu_{2,H}^{(i-1)}\mu_{2,C}^{(i)} \\
&+\frac12 \frac{q_3}{p_3} \tr (\mu_3)^2 + \frac{q_3}{p_3}  \tr \mu_3 \mu_{2,C}^{(1)} + \sum_{i=2}^{a_3-1}c_3^{(i)}\tr \mu_{3,H}^{(i-1)}\mu_{3,C}^{(i)}.
\end{aligned}
\end{equation}
Using $\tr(\mu_1)^2=\tr(\mu_2)^2=\tr(\mu_3)^2$, 
we find that the superpotential simplifies further when \begin{equation}
\frac{q_1}{p_1}+\frac{q_2}{p_2}+\frac{q_3}{p_3}=0\label{!!!}
\end{equation} to \begin{equation}
\begin{aligned}
W= &  \frac{q_1}{p_1}  \tr \mu_1 \mu_{1,C}^{(1)} + \sum_{i=2}^{a_1-1}c_1^{(i)}\tr \mu_{1,H}^{(i-1)}\mu_{1,C}^{(i)} \\
&+ \frac{q_2}{p_2}  \tr \mu_2 \mu_{2,C}^{(1)} + \sum_{i=2}^{a_2-1}c_2^{(i)}\tr \mu_{2,H}^{(i-1)}\mu_{2,C}^{(i)} \\
&+ \frac{q_3}{p_3}  \tr \mu_3 \mu_{2,C}^{(1)} + \sum_{i=2}^{a_3-1}c_3^{(i)}\tr \mu_{3,H}^{(i-1)}\mu_{3,C}^{(i)}.
\end{aligned}\label{vvv}
\end{equation}

We now consider  the R-charge $J_{34}$ of the \Nequals4 system
of one $T_N$ theory and many copies of $T(SU(N))$ theory
\emph{before} the coupling  to \Nequals3 Chern--Simons multiplets.
This assigns charge $+2$ to the moment map fields on the Higgs branch side,
and to $-2$ to those on the Coulomb branch side.
Our final superpotential \eqref{vvv} when the condition \eqref{!!!} is satisfied
preserves $J_{34}$,
and therefore the supersymmetry of the gauged theory enhances to \Nequals4 
when the condition  \eqref{!!!} is satisfied.

It is clear that this analysis can be extended to the 3d theories associated to Seifert manifolds of type $(g;q_i/p_i)$.
All what is needed is to replace the $T_N$ theory with the $T_{N,m,g}$ theory,
which has moment map fields $\mu_i$ for $i=1,\ldots,m$ such that $\tr (\mu_i)^2$ is independent of $i$.
Then, the effective superpotential after eliminating adjoint chiral superfields 
preserves $J_{34}$ if and only if $\sum q_i/p_i=0$,
showing \Nequals4 enhancements.
This is in accord with the geometric criterion we quoted in Sec.~\ref{sub:number-seifert}.

\subsection{Supersymmetry enhancements unaccounted for by holonomy}

Let now briefly come back to the theories we discussed in Sec.~\ref{sec:generalizations},
shown in Fig.~\ref{fig:genus2} and in Fig.~\ref{fig:general}.
These theories are obtained by taking copies of 3d $T_N$ theories
and coupling them via \Nequals3 Chern--Simons multiplets.

According to our discussion in Sec.~\ref{sec:associated},
such theories come from the compactification of $N$ M5-branes 
on a 3-manifold,
since the $T_N$ theory comes from $N$ M5-branes on $S^1$ times a sphere with three holes,
and coupling them via \Nequals3 Chern--Simons multiplets
can be realized by gluing them via appropriate $T$ transformations.
For example, the theory shown in Fig.~\ref{fig:genus2} comes from the geometry
shown in Fig.~\ref{fig:genus2geom}.
Here it would be useful to recall our convention for the $SL(2,\bZ)$ transformation.
On the $i$-th  boundary, we call $F$ the $S^1$ fiber and $C$ the boundary on the base surface.
Let $(C,F)$ be these 1-cycles on the left hand side
and $(C',F')$ be defined analogously on the right hand side.
Then we have \begin{equation}
(C,F)g=(C',F').\label{foofoo}
\end{equation}

\def\TUBE#1#2{
\draw[white,line width=13pt] (#1{}2,#2) to  (#1.6,#2);
\draw  ($( #1{}2,#2)+(0,0.2)$) to  ($( #1.6,#2)+(0,0.2)$);
\draw ($( #1{}2,#2)+(0,-.2)$)  to  ($( #1.6,#2)+(0,-.2)$) ;
\node[circ,inner sep=5pt] (X) at (#1.6,#2) {};
}
\def\HALF#1#2{
\node[circ,inner sep=5pt] (X) at (#1.6,2) {};
\node (X) at (#1.6,1.5){$\times$};
\node (X) at (#1#2,2.3){$S^1$};
\node[draw,ellipse,minimum width=1.8cm,minimum height=3.6cm] (X) at (#1{}2,0) {};
\TUBE{#1}{1}
\TUBE{#1}{0}
\TUBE{#1}{-1}
}
\def\FIG#1#2#3{
\begin{tikzpicture}[scale=1.2]
\HALF{-}{.8}\HALF{+}{.95}
\draw[<->,thick,red] (-.3,1) -- node[above]{$#1$} (.3,1);
\draw[<->,thick,red] (-.3,0) -- node[above]{$#2$} (.3,0);
\draw[<->,thick,red] (-.3,-1) -- node[above]{$#3$} (.3,-1);
\end{tikzpicture}
}
\begin{figure}
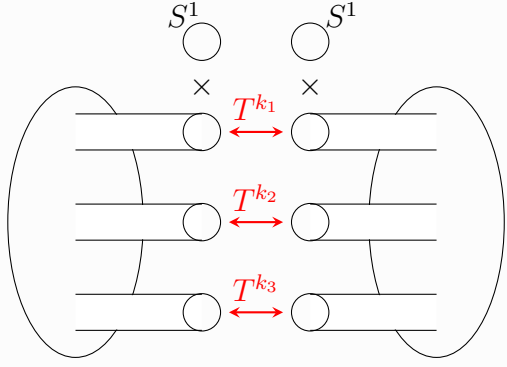

\centering
\FIG{T^{k_1}}{T^{k_2}}{T^{k_3}}
\caption{A geometry generating the theory given in Fig.~\ref{fig:genus2}. 
\label{fig:genus2geom}}
\end{figure}

We saw in Sec.~\ref{sec:generalizations} that these theories have an enhanced \Nequals4 supersymmetry
when 
\begin{equation}
\frac1{k_1}+\frac1{k_2}+\frac1{k_3}=0\label{000}
\end{equation}
The holonomy of the manifold stays the generic $SO(3)$
even when the condition \eqref{000} is satisfied. An argument for this is as follows.\footnote{The authors are grateful to Bruno Martelli for providing to us the arguments in this paragraph and the next.} 
First of all we recall that $SO(2)$ holonomy is equivalent to the existence of a covariantly constant vector field; by \cite[Th.~3]{welsh}, a compact manifold with such a vector is covered by $S^1\times \Sigma$. But now we argue that a graph manifold which is not Seifert cannot be covered by a Seifert (such as $S^1\times \Sigma$). Given a covering $\tilde M \to M$, if $\tilde M$ is modelled on one of the eight Thurston geometries (mentioned at the beginning of this section), then $M$ is also modelled on the same geometry \cite[Cor.~12.9.5]{martelli-book}.
In particular, if $\tilde M=S^1\times \Sigma$ covers a manifold $M$, then $M$ has the same geometry of $S^1\times \Sigma$, and so in particular it is Seifert. 

What is left to be shown is that our graph manifold, shown in Fig.~\ref{fig:genus2geom}, is not a Seifert manifold in disguise.
For Seifert manifolds, every essential surface is either horizontal or vertical with respect to the fibration \cite[Proposition 10.4.9]{martelli-book}.
Now, each of the three $T^2$ where we glue two parts is an essential torus.
If the whole manifold were a Seifert, each of these tori would be either horizontal or vertical with respect to the fibration.
If it is vertical, then the two fibrations on the two pieces glue well along these $T^2$,
which is true only when $k_1=k_2=k_3=0$.
If it is horizontal, the fibers need to lie in intervals in the two pieces.
But this is impossible, because if it is horizontal then its complement is a simple $\mathbb{R}$-bundle, which is not the case.

As a consequence, our graph manifold has $SO(3)$ holonomy; geometrically we would then only expect \Nequals2 supersymmetry, 
leaving the supersymmetry enhancement unaccounted for.

\begin{figure}
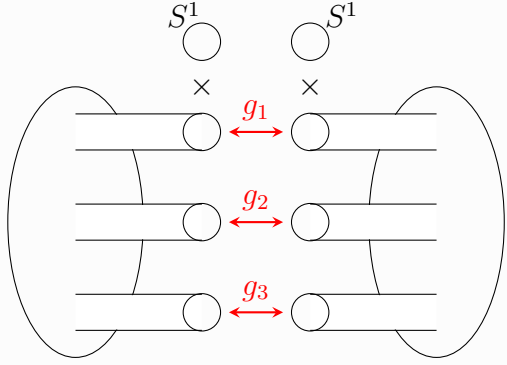

\centering
\FIG{g_1}{g_2}{g_3}
\caption{A further generalization of  Fig.~\ref{fig:genus2geom}. \label{fig:genus2g}}
\end{figure}

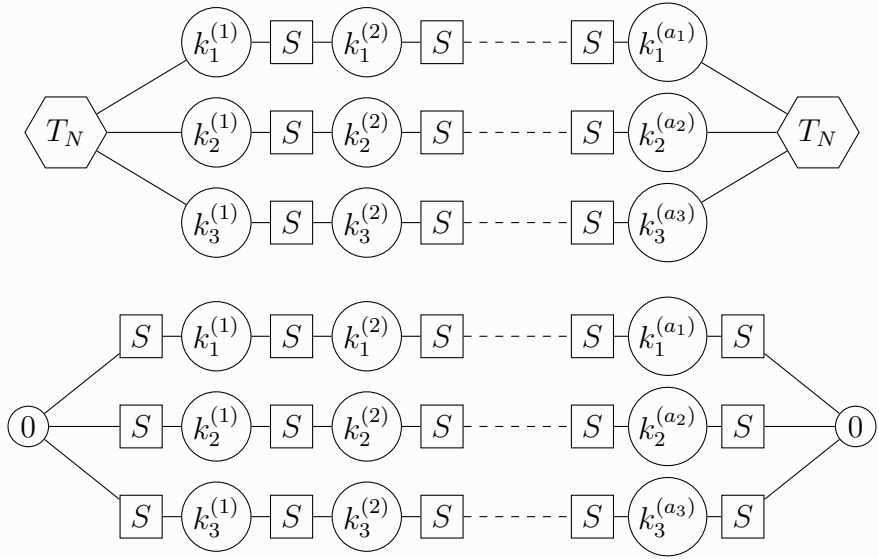
\begin{figure}
\centering
\begin{tikzpicture}
\node[hexagon] (TN) at (0,0) {$T_N$};
\LINE{1.2}{A}1{TN}
\LINE{0}{B}2{TN}
\LINE{-1.2}{C}3{TN}
\node[hexagon] (TNX) at (10,0) {$T_N$};
\draw (A)--(TNX);
\draw (B)--(TNX);
\draw (C)--(TNX);
\end{tikzpicture}

\bigskip

\begin{tikzpicture}
\node[circ,inner sep=2pt] (X) at (-0.5,0) {$0$};
\node[draw,rectangle] (AA) at (1,1.2) {$S$};
\node[draw,rectangle] (BB) at (1,0) {$S$};
\node[draw,rectangle] (CC) at (1,-1.2) {$S$};
\draw (X)--(AA);
\draw (X)--(BB);
\draw (X)--(CC);
\LINE{1.2}{A}1{AA}
\LINE{0}{B}2{BB}
\LINE{-1.2}{C}3{CC}
\node[draw,rectangle] (P) at (9,1.2) {$S$};
\node[draw,rectangle] (Q) at (9,0) {$S$};
\node[draw,rectangle] (R) at (9,-1.2) {$S$};
\node[circ,inner sep=2pt] (Y) at (10.5,0) {$0$};
\draw (A)--(P)--(Y);
\draw (B)--(Q)--(Y);
\draw (C)--(R)--(Y);
\end{tikzpicture}

\caption{The 3d theory associated to the 3-manifold given in Fig.~\ref{fig:genus2g}. 
The second one is obtained by applying the 3d mirror symmetries to the two $T_N$ theories involved.
\label{fig:FIG}}
\end{figure}

We can make a further generalization by using a general $SL(2,\bZ)$ element $g_i$ instead of $T^{k_i}$ in Fig.~\ref{fig:genus2geom}, as shown in Fig.~\ref{fig:genus2g},
whose field-theoretical realization is given in Fig.~\ref{fig:FIG}.
There, we used the two versions, one using two copies of the $T_N$ theory and
another using their 3d mirror descriptions, as discussed in Sec.~\ref{sec:ingredients} and in Fig.~\ref{fig:TN}.

To analyze the enhancement of supersymmetry, we can use the \Nequals2 superpotential \eqref{W} of the duality wall theory realizing an arbitrary $g\in SL(2,\bZ)$.
Writing \begin{equation}
g_i = \begin{pmatrix}
q'_i &  p_i \\
r_i & q_i
\end{pmatrix},
\end{equation} 
and using our convention \eqref{foofoo}, we see that the boundary 1-cycles are pasted 
such that $F'=p_i C + q_i F$ and $F=-p_i C' + q'_i F'$.
It should be by now easy for the reader to see that the supersymmetry enhances when both the following relations
\begin{equation}\label{eq:qp-qpp}
\frac{q_1}{p_1}+\frac{q_2}{p_2}+\frac{q_3}{p_3}=0,
\qquad
\frac{q_1'}{p_1}+\frac{q_2'}{p_2}+\frac{q_3'}{p_3}=0
\end{equation}
are satisfied.
For example, in the second description in Fig.~\ref{fig:FIG},
we use the fact that the duality wall theory implementing $\hat g_i= S T^{k_i^{(1)}} S \cdots S T^{k_i^{(a_i)}}S$ 
has an induced background Chern--Simons terms with the coupling \begin{equation}
\frac12\frac{q}{p} \tr \Phi^2
\end{equation}
for the first $SU(N)$ symmetry.
This follows easily from our analysis in Sec.~\ref{sec:interlude} and the Statement there;
we also have a few more comments on it in Appendix~\ref{app:wall}.
Therefore, the $SU(N)$ gauging on the far left in the second description in Fig.~\ref{fig:FIG} has an induced
\Nequals3 Chern--Simons coupling whose coefficient is $q_1/p_1+q_2/p_2 + q_3/p_3$,
and allows \Nequals4 enhancement if this combination vanishes.

\begin{figure}
\centering
\begin{tikzpicture}
\node[circ,inner sep=1mm,fill] (B1) at (0,0) {};
\node[circ,inner sep=1mm,fill] (B2) at (4,0) {};
\node[circ,inner sep=1mm] (W1) at (2,1) {};
\node[circ,inner sep=1mm] (W2) at (2,-1) {};
\draw (B1)--node[above,red]{$g_e$} (W1);
\draw (B1) edge[bend left] (W2);
\draw (B1) edge[bend right] (W2);
\draw (B2) edge[bend left] (W2);
\draw (B2) edge[bend right] (W2);
\draw (B2) edge[bend left] (W1);
\draw (B2) edge[bend right] (W1);
\end{tikzpicture}
\caption{A further generalization of Fig.~\ref{fig:genus2g}. Every link $e$ represents a coupling of two $T_N$ theories by a superpotential of the form \eqref{W}, corresponding to a $g\in SL(2,\mathbb{Z})$ as in \eqref{g}.  \label{fig:graphman}}
\end{figure}
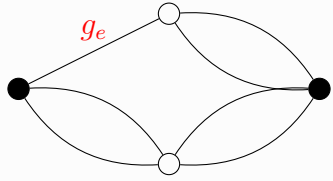

Our final generalization is to use a general bipartite graph whose edge is labeled by $SL(2,\bZ)$ elements, as shown in Fig.~\ref{fig:graphman}.
This class of 3-manifolds (without the bipartite assumption) is known as graph manifolds.
As a 3d theory, it generalizes the theories we discussed Fig.~\ref{fig:general} in Sec.~\ref{sec:generalizations}.
For each edge $e$, define $p_e$, $q_e$, $q'_e$ analogously using the associated $SL(2,\bZ)$ transformation $g_e$.
Then, the supersymmetry enhances if 
at each black node $v$, the sum of $q_e/p_e$ of edges $e$ connecting to $v$ vanishes,
and
at each white node $v$, the sum of $q'_e/p_e$ of edges $e$ connecting to $v$ vanishes.

In Appendix~\ref{sec:homology}, we study the homology groups of these manifolds,
where we find some hints that something is going on when the conditions \eqref{000} or \eqref{eq:qp-qpp} are met.
In Appendix~\ref{sec:sugra}, we study a more general supergravity backgrounds on these manifolds,
and again we find some hints that something is going on on this class of manifolds.
But in neither Appendices we could find definitive evidence from geometry
that the supersymmetry would enhance.

There are other cases of field-theoretical \Nequals4 enhancements
which are unaccounted for from geometry.
As in Sec.~\ref{sec:mainexamples}, consider the $T_{N,n}$ theory whose $i$-th $SU(N)$ flavor symmetry is gauged with the Chern-Simons level $k_i$.
We now allow some of $k_i$ to be zero.  
The field theoretical analysis can be repeated, and we find that the supersymmetry enhances to \Nequals4 if \begin{equation}
\sum_{k_i\neq 0} \frac{1}{k_i}=0.
\end{equation}
The corresponding 3d manifolds can still be obtained by Dehn fillings,
but are not Seifert manifolds, which do not allow $k_i=0$.
The mathematical theorem quoted in Sec.~\ref{sub:number-seifert} does not apply,
and therefore the \Nequals4 enhancments are unaccounted for geometrically.

Finally, we should mention that  there is a bigger problem behind these \Nequals4 enhancements, in a certain sense.
From the construction explained in Sec.~\ref{sec:associated},
it is clear that all theories obtained from gluing Seifert manifolds along $T^2$ boundaries
have \Nequals3 supersymmetry,
since they are obtained by combining copies of \Nequals4 $T_N$ theories, 
and performing $SL(2,\bZ)$ transformations, 
which uses only copies of \Nequals4 $T(SU(N))$ theories
and \Nequals3 Chern--Simons multiplets. 
That said, such geometries would have $SO(3)$ holonomy,
and we only expect \Nequals2 supersymmetry in these cases.
Therefore, this generic enhancement to \Nequals3 is similarly unaccounted for by holonomy.

\section*{Acknowledgments}
The authors thank Jin-Mann Wong and Seyed Morteza Hosseini for the collaboration during the early stages of the work, and Bruno Martelli, Antoine Van Proeyen, Alberto Zaffaroni for discussions.
The authors also thank the organizers (Kazunobu Maruyoshi and Jaewon Song) of the workshop ``\href{https://sites.google.com/view/grtquantumfields/}{Geometry, Representation Theory and Quantum Fields}'' held virtually in Osaka this year
for giving them an ideal opportunity to revive this project, which had been untouched for more than a year at that point.
YT is supported in part  
by WPI Initiative, MEXT, Japan at Kavli IPMU, the University of Tokyo
and by JSPS KAKENHI Grant-in-Aid (Kiban-S), No.16H06335.
AT is supported in part by INFN and by MIUR-PRIN contract 2017CC72MK003.

\appendix

\section{Homology} 
\label{sec:homology}

Here we will compute the dimension of the homology of the spaces we consider in the main text, showing that it displays jumping phenomena similar to the ones we found from a physics point of view. While we are not aware of a direct connection between the two, we think this parallel behavior is rather suggestive and might have a deeper significance.  

In general, if we write our space $M$ as a union of two $X_1$ and $X_2$, the homology groups are related by the \emph{Mayer--Vietoris} exact sequence
\begin{equation}\label{eq:mayer-vietoris}
	\ldots \to H_{k+1}(M) \to H_k(X_1 \cap X_2) \to H_k(X_1) \oplus H_k(X_2) \to H_k(M)\to \ldots\,.
\end{equation}
As usual, this means that the image of each map is equal to the kernel of the one that follows. The first map in this sequence is a restriction, the others are natural inclusions. As we will see in the examples, the intuitive understanding is that every cycle in $H_k(M)$ is either realized by gluing two cycles in $X_1$ and $X_2$, or comes from one on $X_1 \cap X_2$.

\paragraph{Seifert manifolds:}
Our first application is to Seifert manifolds. Here we actually know already from section \ref{sub:number-seifert} that the topology changes in correspondence to the enhancement predicted by field theory; we include this discussion as a warm-up. For simplicity we consider the manifold corresponding to the theory in Fig.~\ref{fig:gaugedTn}. Following our description in section \ref{sec:moreexamples}, this is obtained from $S^1\times S^2$ by cutting three $S^1 \times D^2_i$ and gluing them to $S^1\times (D^2)'_i$  after a $T^{k_i} \in SL(2,\bZ)$ transformation. In the convention of (\ref{eq:peculiar}),
\begin{equation}\label{eq:STki}
	C_i= C'_i\, ,\qquad F_i= F'_i-k_i C'_i
\end{equation} 
(with no sum over $i$). Now take $X_1= S^1\times (S^2- \cup_i D^2_i)$, and $X_2 = S^1\times \cup_i (D^2)'_i$. We can compute some of the homology groups in (\ref{eq:mayer-vietoris}) right away: 
\begin{itemize}
	\item The homology groups of $X_1$ are: $H_0=\mathbb{Z}$; $H_1=\mathbb{Z}^3$ (with generators $C_1$, $C_2$, $F$); $H_2(X_1)= \mathbb{Z}^2$ (with generators $C_1\times F$, $C_2 \times F$). Notice that the boundary $C_3= \partial D^2_3$ of the third disk is homologous to $-C_1- C_2$ on $X_1$, and that the $F_i$ are all in the same homology class, which we simply call $F$. 
	\item Each $S^1 \times(D^2)'_i $ has $H_0= \mathbb{Z}$, $H_1=\mathbb{Z}$ (generated by the fiber $C'_i$), and $H_2=0$.
	\item Finally, $X_1 \cap X_2$ consists of three copies of $T^2$, each of which has $H_0=\mathbb{Z}$, $H_1=\mathbb{Z}^2$ (whose generators we call $b_i$, $f_i$), $H_2 = \mathbb{Z}$ (generated by $f_i\times b_i$, no sum).
\end{itemize}
The sequence (\ref{eq:mayer-vietoris}) now reads
\begin{equation}
	0 \to H_3(M) \to \mathbb{Z}^3 \to \mathbb{Z}^2 \to H_2(M)\to \mathbb{Z}^6 \buildrel{\phi}\over \to \mathbb{Z}^6 \to H_1(M) \to \mathbb{Z}^3 \to \mathbb{Z}^4 \to H_0(M)\to 0\,.
\end{equation}
Using exactness of the sequence at every step and knowing $H_0(M)=H_3(M)=\mathbb{Z}$, we find
\begin{equation}\label{eq:hi-seifert}
	h_1(M)= h_2(M)=\mathrm{dim}(\mathrm{ker}\phi)\,,
\end{equation}
where $h_k=\mathrm{dim}(H_k)$. (As we said at the beginning, we focus on the dimension over $\mathbb{Z}$, ignoring possible ``torsion'' terms, summands of the type $\mathbb{Z}_m$.)

The map $f: H_1(X_1\cap X_2)\to H_1(X_1) \oplus H_1(X_2)$ can be found as follows. We take $H_1(X_1\cap X_2)\to H_1(X_1)$ to be induced by the natural inclusion $X_1\cap X_2\subset X_1$: so the $f_i$ are all mapped on $X_1$ to the $F_i\sim F$, and the $b_i$ are mapped to $\partial D^2_1 = C_1$, $\partial D^2_2 =C_2$, $ \partial D^2_3 = -C_1-C_2$. We now recall (\ref{eq:STki}); since the disk boundaries $F'_i$ are in fact trivial in $X_2$, we have $f_i = k_i C'_i$, $b_i= - C'_i$. All in all this gives the matrix 
\begin{equation}\label{eq:phi-seifert}
	\phi= \left(\begin{array}{cccccc}
		1 & 0 & 1 & 0 & 1 & 0 \\
		0 & 1 & 0 & 0 & 0 & -1\\
		0 & 0 & 0 & 1 & 0 & -1\\
	   k_1&-1 & 0 & 0 & 0 & 0 \\
		0 & 0 &k_2&-1 & 0 & 0 \\
		0 & 0 & 0 & 0 &k_3&-1 
	\end{array}\right)\,.
\end{equation}
\begin{itemize}
	\item For generic $k_i$, $\phi$ is non-degenerate, so (\ref{eq:hi-seifert}) gives $h_1(M)=h_2(M)=0$.
	\item When the condition $\sum_i 1/k_i=0$ is satisfied, the vector $(-1/k_1,1,-1/k_2,1,-1/k_3,1)$ is in the kernel and $\phi$ has rank 5. So $h_1(M)=h_2(M)=1$. 
\end{itemize}
As expected, the homology dimensions jump when (\ref{yyy}) is satisfied. 

We can generalize this example replacing $T^{k_i}$ by arbitrary elements $g_i$; in the convention of (\ref{g}), the $(k_i,-1)$ in the lower half of (\ref{eq:phi-seifert}) are replaced by $(p_i,-q_i)$. Again $\phi$ is generically non-degenerate, but has rank 5 when $\sum_i q_i/p_i=0$, with a kernel spanned by 
\begin{equation}\label{eq:ker-seifert}
	\left(\frac{q_1}{p_1},1,\frac{q_2}{p_2},1,\frac{q_3}{p_3},1\right)\,;
\end{equation}
in the latter case $h_1(M)=h_2(M)=1$. 

The further generalization to a larger number $K$ of singular fibers is straightforward; while the determinant becomes harder to compute, the generalization $(q_1/p_1,1,\ldots,q_K/p_K,1)$ of (\ref{eq:ker-seifert}) is still correct.

\paragraph{Two-node graph manifold:}
We now consider the case of graph manifolds. This is more interesting, in the sense that here we have not yet identified a geometrical counterpart of enhancement. 

We again begin with an example: the case in Fig.~\ref{fig:genus2geom}. Again $X_1= S^1\times (S^2- \cup_i D^2_i)$, but now we glue it to a second copy $X_2= S^1\times (S^2- \cup_i (D^2)'_i)$. On each tube, the identification is given by (\ref{foofoo}) with $g_i = T^{k_i}$:
\begin{equation}\label{eq:2node-STki}
	F'_i= F_i + k_i C_i \, ,\qquad C'_i=C_i\,.
\end{equation} 

The homology groups of $X_1$ and $X_2$ are the same as that of $X_1$ in the previous Seifert example. Also $X_1 \cap X_2$ is the same as in the previous example. (\ref{eq:mayer-vietoris}) now reads
\begin{equation}
	0 \to H_3(M) \to \mathbb{Z}^3 \to \mathbb{Z}^4 \to H_2(M)\to \mathbb{Z}^6 \buildrel{\phi}\over \to \mathbb{Z}^6 \to H_1(M) \to \mathbb{Z}^3 \to \mathbb{Z}^2 \to H_0(M)\to 0\,.
\end{equation}
Exactness of the sequence now implies
\begin{equation}
	h_1(M)= h_2(M)= 2 + \mathrm{dim}(\mathrm{ker}\phi)\,.
\end{equation}
Intuitively, the 2 one-cycles in $h_1$ correspond to the image of the map $H_1(M)\to H_0(X_1\cap X_2)=\mathbb{Z}^3$: in other words, these cycles intersect $X_1\cap X_2$. When $k_i=0$ and $X=S^1 \times \Sigma_{g=2}$, these are the A-cycles of the genus-2 Riemann surface $\Sigma_{g=2}$. The additional $\mathrm{dim}(\mathrm{ker}\phi)$ one-cycles correspond to the image of the map $H_1(X_1)\oplus H_1(X_2)= \mathbb{Z}^6 \to H_1(M)$: they come from one-cycles of the two halves. When $k_i=0$, these are the B-cycles of $\Sigma_{g=2}$, plus the $S^1$ fiber. 

To find $\phi$, we can embed the generators of $H_1(X_1\cap X_2)$ into those of $H_1(M)$ in the same way as in the previous Seifert case; so the upper half of the matrix should be identical to those of (\ref{eq:phi-seifert}). But the way we embed them in $H_2(M)$ is now dictated by (\ref{eq:2node-STki}), leading to 
\begin{equation}\label{eq:phi-2node}
	\phi= \left(\begin{array}{cccccc}
		1 & 0 & 1 & 0 & 1 & 0 \\
		0 & 1 & 0 & 0 & 0 & -1\\
		0 & 0 & 0 & 1 & 0 & -1\\
	    1 & 0 & 1 & 0 & 1 & 0 \\
	   k_1& 1 & 0 & 0 &-k_3&-1 \\
		0 & 0 &k_2& 1 &-k_3&-1 
	\end{array}\right)\,.
\end{equation}
\begin{itemize}
	\item If the $k_i$ are generic, $\mathrm{dim}(\mathrm{ker}\phi)=1$ (generated by $(0,1,0,1,0,1)$), and $h_1=h_2=3$. 
	\item If $\sum_i 1/k_i=0$, $\mathrm{dim}(\mathrm{ker}\phi)=2$ (with additional generator $(1/k_1,0,1/k_2,0,1/k_3,0)$), and $h_1=h_2=4$.
	\item If $k_i=0$, $\mathrm{dim}(\mathrm{ker}\phi)=3$, and $h_1=h_2=5$; this is the $S^1\times \Sigma_{g=2}$ case we mentioned earlier.
\end{itemize}
Thus we see an enhancement of homology when (\ref{000}) is satisfied.

If we generalize the $T^{k_i}$ on each node to general $g_i \in SL(2,\mathbb{Z})$, we obtain the manifold in Fig.~\ref{fig:genus2g}. The relevant matrix is
\begin{equation}\label{eq:phi-2node-gen}
	\phi= \left(\begin{array}{cccccc}
		1 & 0 & 1 & 0 & 1 & 0 \\
		0 & 1 & 0 & 0 & 0 & -1\\
		0 & 0 & 0 & 1 & 0 & -1\\
	   q'_1&-r_1&q'_2&-r_2&q'_3&-r_3\\
	  -p_1&q_1& 0 & 0 &p_3&-q_3 \\
		0& 0 &-p_2&q_2&p_3&-q_3
	\end{array}\right)\,.
\end{equation}
Define now the vectors 
\begin{equation}
	v_1 \equiv \left(\frac1{p_1},0,\frac1{p_2},0,\frac1{p_3},0\right) \, ,\qquad
	v_2 \equiv \left(\frac{q_1}{p_1},1,\frac{q_2}{p_2},1,\frac{q_3}{p_3},1\right)\,.
\end{equation}
It is easy to see that $\phi v_1= (\sum_i 1/p_i,0,0,\sum_i q'_i/p_i,0,0)$, and $\phi v_2 = (\sum_i q_i/p_i,0,0,\sum_i 1/p_i,0,0)$. 
\begin{itemize}
	\item When the $g_i$ are totally generic, $\phi$ is non-degenerate: $h_1=h_2=2$.
	\item When $(\sum_i \frac{q_i}{p_i})(\sum_i \frac{q'_i}{p_i})= (\sum_i \frac1{p_i})^2$, the vector $(\sum_i q_i/p_i) v_1 - (\sum_i 1/p_i )v_2$ is in the kernel, so $h_1=h_2=3$.
	\item When $\sum_i 1/p_i= \sum_i q_i/p_i=\sum_i q'_i/p_i=0$, both $v_1$ and $v_2$ are in the kernel; so $h_1=h_2=4$.
\end{itemize}
So in this case the correspondence with the field theory analysis isn't quite as precise, but the condition (\ref{eq:qp-qpp}) does appear.

    \paragraph{Torus bundles:}
As a cross-check, we can consider a graph with two nodes and two links rather than three. We obtain a torus bundle over $S^1$. If we view it as $(T^2\times [0,1])/\sim$ with an identification $(t,0)\sim (g.t,1)$, $t\in T^2$, then $g\in SL(2,\mathbb{Z})$ is related to the $g_1$ and $g_2$ on the two tubes by $g=g_1 g_2^{-1}$. 
We will also consider these spaces in section \ref{sssec:torusbundle} and at the end of section \ref{ssub:torus-bundles}. 

The computation in this case is almost identical as in the previous one. We find $h_1(M)=h_2(M)=1+\mathrm{dim}(\mathrm{ker}\phi)$; we find it easier to use a slightly different convention for the $C_i$, such that $\phi= \left(\begin{smallmatrix}
1_2 & 1_2 \\ g_1^{-1} & g_2^{-1}
\end{smallmatrix}\right)$.
Now $\det \phi = 2-\mathrm{Tr}(g_1 g_2^{-1})$. 
\begin{itemize}
	\item If $\mathrm{Tr}(g)>2$, $\det \phi\neq 0$, and $h_1=1$. Indeed this space is known as Sol: it is a discrete quotient of a solvable group, whose Lie algebra can be summarized by the Maurer--Cartan relations $\mathrm{d} e_1 = - e_1 \wedge e_3$, $\mathrm{d} e_2 = e_2 \wedge e_3$, $\mathrm{d} e_3=0$. Among these left-invariant one-forms, only $e_3$ is in cohomology. By \cite[Thm. 3.11b]{bock}, the cohomology computed on left-invariant forms is the same as the de Rham cohomology, so our computation is correct.
	\item If $\mathrm{Tr}(g)=2$, $\mathrm{dim}(\mathrm{ker}\phi)= 1$, and $h_1=2$. Indeed this is an $S^1$-bundle over $T^2$, so the two one-forms of the base are in cohomology, that of the fiber is not.
 	\item If $\mathrm{Tr}(g)<2$, $\det \phi\neq 0$, and $h_1=1$. This is a quotient of the solvable group with $\mathrm{d} e_1 = e_2 \wedge e_3$, $\mathrm{d} e_2 = -e_1 \wedge e_3$, $\mathrm{d} e_3=0$; again only $e_3$ is in cohomology.
\end{itemize}

\paragraph{General graph manifolds:}

Consider now a graph manifold based on a general bipartite graph,  as in Fig.~\ref{fig:general} and \ref{fig:graphman}. Here we can take $X_1$ to be the union of the multi-punctured spheres corresponding to all black nodes, and $X_2$ of the spheres corresponding to white nodes; let $v$ be the total number of nodes. $X_1\cap X_2$ will be a union of $e= \#(\text{edges})$ copies of $S^1\times T^2$.  (\ref{eq:mayer-vietoris}) is now
\begin{equation}
	0 \to H_3(M) \to \mathbb{Z}^{e} \to \mathbb{Z}^{2e-v} \to H_2(M)\to \mathbb{Z}^{2e} \buildrel{\phi}\over \to \mathbb{Z}^{2e} \to H_1(M) \to \mathbb{Z}^e \to \mathbb{Z}^v \to H_0(M)\to 0\,.
\end{equation}
We obtain $h_1(M)=h_2(M)= e-v+1 + \mathrm{dim}(\mathrm{ker}\phi)$. The properties of $\phi$ are similar to those of (\ref{eq:phi-2node}), (\ref{eq:phi-2node-gen}) for the two-node graph. For example, when we have $T^{k_i}$ transformations on the links as in Fig.~\ref{fig:general}: 
\begin{itemize}
	\item For generic $k_i$, $\mathrm{dim}(\mathrm{ker}\phi)=1$, with a single generator $(0,1,\ldots,0,1)$. 
	\item When the condition $\sum_i 1/k_i=0$ is satisfied at every node, $\mathrm{dim}(\mathrm{ker}\phi)=2$, with an additional generator $(1/k_1,0,\ldots,1/k_e,0)$.
\end{itemize}

\paragraph{Some connections to topological properties of graph manifolds:}

The above results show that the homology groups are enlarged when certain conditions are met, such as $\sum_i 1/k_i = 0$. 
The conditions found do not match in all cases the ones found for \Nequals{4} enhancement of the associated field theory, however
they seem related to some theorems in \cite{Neumann,BuyaloSvetlov}, which study topologies of graph manifolds
(in particular Theorem 4.1 in \cite{Neumann} and Theorem 4.7 in \cite{BuyaloSvetlov}).

In those works, one constructs an $n\times n$ matrix $S$ associated to a graph manifold, where $n$ is the number of nodes in the graph. 
In a nutshell, the diagonal elements of S are (minus) the effective Chern-Simons levels at each node 
$- k_v = - \sum_{\text{$e$:meeting at $v$}} \frac{q_e}{p_e}$, after removing the complex 
chirals $\Phi_i$ as in the main text, and the off-diagonal elements are $\sum_{\text{$e$:connecting $(v_1,v_2)$}} \frac1p_e$.

One theorem says that the manifold admits a horizontal surface (i.e.~an embedded surface transverse to a Seifert fiber on a Seifert component of the manifold)
if and only if $\det(S)=0$. This matches some conditions found above for 
homology group enlargements, e.g.~the condition $(\sum_i \frac{q_i}{p_i})(\sum_i \frac{q'_i}{p_i})= (\sum_i \frac1{p_i})^2$ in 
the 2-node case.
Another theorem says that the manifold is a surface bundle over an $S^1$ circle if and only if ker$(S)$ contains a vector with all elements non-zero.
We have not been able to relate this condition to a property of the gauge theory.
It would be desirable to study the connection between these geometric properties and gauge theory properties further. 


\section{Background supergravity} 
\label{sec:sugra}

Since we have no evidence of special holonomy on these spaces, it would be interesting to find if there is any other geometrical mechanism for the enhancement we see field-theoretically. 

\paragraph{Conformal ${\mathcal N}=(2,0)$ supergravity:}
To see what other geometric structure might be at play, consider the 6d ${\mathcal N}=(2,0)$ theory on the space $\mathrm{Mink}_3 \times M$. A standard approach \cite{Festuccia:2011ws} is to couple the theory to background conformal ${\mathcal N}=(2,0)$ supergravity \cite{Bergshoeff:1999db}.\footnote{See \cite{Samtleben:2012ua} for an analysis of this approach in six dimensions.} To see how much supersymmetry is preserved, we then have to solve the transformations of the fermionic fields. For the gravitino:
\begin{equation}\label{eq:gravitino}
	\delta \psi^i_\mu = {\mathcal D}_\mu \epsilon^i - \frac14 T^i{}_j\gamma_\mu \epsilon^j + \gamma_\mu \tilde \epsilon^i\,.
\end{equation}
Here ${\mathcal D}_\mu$ is the derivative covariant with respect to both the spin connection and the  $USp(4)_R$ gauge field $V^i{}_{j\mu}$, $i=1,\,\ldots,\,4$; $T^{ij}=\frac16 T^{ij}_{\mu \nu \rho} \gamma^{\mu \nu \rho}$, with $T^{ij}_{\mu \nu \rho}$ self-dual; $\epsilon^i$, $\tilde \epsilon^i$ are the parameters for the $Q$ and $S$ transformations respectively, are chiral and satisfy a symplectic Majorana condition, $(\epsilon^i)^\mathrm{c}= \epsilon^j \Omega_{ji}$, with $\Omega= \mathrm{i} 1_2 \otimes  \sigma_2$. 

\paragraph{$SO(2)$ holonomy:}
In the case with $SO(2)$ holonomy, we can take the gauge field to be along an $so(2)$ subalgebra, and $\epsilon^i$ an eigenvalue of its action: for example $V^i{}_{j\mu}= A_\mu (-\mathrm{i} \sigma_2 \otimes  1_2)^i{}_j$, and $\epsilon^i= (\zeta_1,\zeta_2, \mathrm{i} \zeta_1, \mathrm{i} \zeta_2)^i$,  so that $(V_\mu \epsilon)^i= \mathrm{i} A_\mu \epsilon^i$. 
Moreover we take $V$ to be along $M$. Now we further decompose along the three external and three internal directions: $\gamma_\alpha = \gamma^{(3)}_\alpha \otimes 1 \otimes \sigma_1$, $\gamma_m = 1 \otimes \sigma_m \otimes \sigma_2$. For each 6d spinor we take the Ansatz 
$\zeta= \xi \otimes \eta \otimes {1\choose 0} + \mathrm{c.c.}$, 
with the last factor taking care of chirality. With all this, one can check that (\ref{eq:gravitino}) reduces to 
\begin{equation}\label{eq:twist}
	D^A_m \eta =0 \, ,\qquad D^A_m\equiv D_m - \mathrm{i} A_m
\end{equation}
on $M$. The transformations for the other fermionic field, $\delta \chi^{ij}_k$, are automatically zero.  A covariantly constant spinor, $D_m \eta=0$, would reduce holonomy to the stabilizer of $\eta$, which is just the identity. But the presence of the $so(2)$ connection means that a spinor obeying (\ref{eq:twist}) reduces holonomy to $so(2)$; this is usually called a \emph{twist}. 

So our Ansatz solves the supersymmetry equations on any $M$ of reduced holonomy. Moreover, given that $\xi$ can be any Dirac spinor on Mink$_3$ and that we had two possible $\zeta$, we are preserving a total of eight supercharges, or ${\mathcal N}=4$, as expected.

As a simple example, consider the $T^2$-bundles,
which we also discuss in \ref{ssub:torus-bundles}. They are all obtained as $(T^2\times \bR)/\sim$, with the relation $(t,z)\sim (g.t,z+1)$. When $g$ is elliptic, it is a rotation by an angle $\alpha$. A chiral spinor $\eta_0$ on $T^2$ can be promoted to a spinor on $T^2\times \bR$ as usual by viewing the chiral $\gamma= -\mathrm{i} \gamma_{12}$ as $\gamma_z$. But the rotation means that the frame before and after one turn in the $z$ direction do not coincide; correspondingly, the spinor transforms by $\eta_0\to \mathrm{e}^{\alpha \gamma_{12}/2} \eta_0=\mathrm{e}^{\mathrm{i}\alpha /2} \eta_0$. But the spinor $\eta= \mathrm{e}^{-\mathrm{i}\alpha z/2} \eta_0$ is then well-defined, and satisfies (\ref{eq:twist}) with $A=-\alpha/2 \mathrm{d}z$.

\paragraph{Looking for general solutions:}
A more general analysis is more daunting, but we note here an interesting feature. Take a warped product $\mathrm{d}s^2_6= \mathrm{e}^{2W} \mathrm{d}s^2_{\mathrm{Mink}_3} + \mathrm{d}s^2_{M}$; Minkowski symmetry dictates to again take $D_\mu \eta$ and $V_\mu$ only along $M$, and $T^i{}_j = t^i{}_j(\mathrm{vol}_{\mathrm{Mink}_3} - \mathrm{vol}_{M})$ in form notation. (\ref{eq:gravitino}) now implies
\begin{subequations}\label{eq:dpsi-dec}
\begin{align}
	\label{eq:dpsi-ext} (T^\dagger)^i{}_j \epsilon^j + \left(D^V +\frac32 \partial W\right) \epsilon = 0 \,,\\
	\label{eq:cks}\left(D^V_m -\frac13 \gamma_m D^V\right) \epsilon =0 \,.
\end{align}
\end{subequations}
where $D^V= \gamma^m D^V_m$, $\partial W = \gamma^m \partial_m W $. 

If we now again factorize $\zeta= \xi \otimes \eta \otimes {1\choose 0} + \mathrm{c.c.}$, take $V$ abelian and $(V_\mu \epsilon)^i= \mathrm{i} A_\mu \epsilon^i$, (\ref{eq:cks}) becomes the (charged) \emph{conformal Killing spinor} (CKS) equation $(D^A_m -1/3 \gamma_m D^A) \eta =0 $ on $M$. This was shown in turn \cite{Klare:2012gn} to be equivalent to the existence of a so-called \emph{transversely holomorphic foliation (THF)}: a complex one-form $o$ that is non-degenerate ($o \wedge \bar o \neq 0$ everywhere) and such that 
\begin{equation}
	\mathrm{d} o = w \wedge o
\end{equation}
for some $w$. This condition is formally identical to a possible definition of a complex structure in even dimensions. Indeed there is also an alternative description of a THF in terms of a $(1,1)$-tensor $I_m{}^n$. This structure appears in the study of $d=3$ ${\mathcal N}=2$ theories on curved spaces \cite{Klare:2012gn,Closset:2012ru}, and so in a sense it is quite natural to see it appear as well for compactifications on a three-manifold.

However, the existence of a THF is not enough to guarantee supersymmetry within this approach: both (\ref{eq:dpsi-dec}) and the remaining supersymmetry transformations $\delta \chi^{ij}_k$ give further constraints. Moreover, while Seifert manifolds and torus bundles with hyperbolic $g\in SL(2,\mathbb{Z})$ admit a THF, the rest of the spaces we considered in this paper do not. (THFs are classified in \cite{brunella,ghys}.) So it is likely that we need to go back to (\ref{eq:cks}), which more generally looks like a non-Abelian generalization of the charged CKS equation. We hope to return to this in the future.


\section{More on the \texorpdfstring{$SL(2,\bZ)$}{SL(2,Z)} duality wall theory}
\label{app:wall}

In this appendix, we make a further remark on the $SL(2,\bZ)$ duality wall theory,
using the analysis of the superpotential performed in Sec.~\ref{sec:interlude}.

\subsection{Effective superpotential and the induced  Chern--Simons couplings}
\begin{figure}
\centering
\begin{tikzpicture}
\node[draw,rectangle] (m1) at (1,1.2) {$S$};
\node[circ] (k1) at (2,1.2) {$k^{(1)}$};
\node[draw,rectangle] (n1) at (3,1.2) {$S$};
\node[circ] (k1x) at (4,1.2) {$k^{(2)}$};
\node[draw,rectangle] (n1x) at (5,1.2) {$S$};
\node[draw,rectangle] (n1y) at (7,1.2) {$S$};
\node[circ] (k1y) at (8,1.2) {$k^{(a)}$};
\node[draw,rectangle] (mm) at (9,1.2) {$S$};
\draw (m1)--
(k1)--(n1)--(k1x)--(n1x);
\draw[dashed] (n1x)--(n1y);
\draw (n1y)--(k1y)--(mm);
\end{tikzpicture}
\caption{An $SL(2,\bZ)$ wall theory implementing $g'=ST^{k^{(1)}} S\cdots S T^{k^{(a)}}S$.
 \label{fig:wall}}
\end{figure}
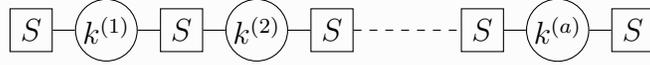

In Sec.~\ref{sec:interlude}, we analyzed a theory associated to Fig.~\ref{fig:chain}.
Here we are more interested in the duality wall theory itself, shown in Fig.~\ref{fig:wall},
obtained by further coupling the $T(SU(N))$ theory on both ends, implementing the transformation \begin{equation}
\hat g=ST^{k^{(1)}} S\cdots S T^{k^{(a)}}S.\label{ghat}
\end{equation}

It is easy to derive the effective superpotential of this theory from the Statement and the Lemma established in Sec.~\ref{sec:interlude};
we have \begin{equation}
W = \frac12 \frac{q}{p} \tr \Phi^2  + \tr\Phi \mu_C^{(0)}
+ \sum_{i=1}^{a}c^{(i)}\tr \mu_H^{(i-1)}\mu_C^{(i)} 
+ \tr\mu_H^{(a)} \Phi'+ \frac12\frac{q'}{p} \tr ( \Phi' )^2.\label{eff}
\end{equation}
where $\Phi$ and $\Phi'$ are the adjoint chiral superfields in the background vector multiplets
coupled to the two $SU(N)$  flavor symmetries.
We can also establish the chiral ring relations \begin{equation}
\tr (\mu_C^{(0)} )^2 = \frac{1}{p^2} \tr (\Phi')^2,\qquad
\tr (\mu_H^{(a)} )^2 = \frac{1}{p^2} \tr (\Phi)^2.
\label{crwall}
\end{equation}

This superpotential preserves the symmetry $J_{34}$, except for the purely background terms
proportional to  $ \tr \Phi^2$ and $ \tr (\Phi')^2$.
Therefore, these $SL(2,\bZ)$ wall theories are \Nequals4 supersymmetric if we do not couple them to $SU(N)$ background fields.

The coefficients of $\tr\Phi^2$ and $\tr(\Phi')^2$ are related by \Nequals3 supersymmetry to the contact terms of the current-current two-point functions, or equivalently the coefficients of the background Chern--Simons terms,
which can be computed by localization on $S^3$ \cite{Closset:2012vg,Closset:2012vp}.
Let us perform this computation, which gives a nice consistency check of our superpotential manipulation in Sec.~\ref{sec:interlude}.

We base our computation on the $S^3$ partition function of the $T(SU(N))$ theory, which has the form \cite{Benvenuti:2011ga,Nishioka:2011dq} \begin{equation}
Z_{T(SU(N))} (s,s') = \sum_{w\in W} \frac{(-1)^w e^{2\pi \mathrm{i} s w(s')} }{\Delta(s)\Delta(s')}
\end{equation} where $s,s'\in \bR^{N-1}$ are the real scalar fields in the background  \Nequals2 vector multiplet of two $SU(N)$ flavor symmetries, $W=\mathfrak{S}_N$ is the Weyl group, and $\Delta(s)$ is the measure factor appearing in the contribution of the vector multiplet, which is \begin{equation}
\frac{1}{N!}\Delta(s)^2 d^{N-1} s.
\end{equation}
The Chern--Simons contribution of level $k$ is in turn given by\begin{equation}
e^{\pi \mathrm{i} s^2}.
\end{equation}
Then the partition function of the duality wall theory implementing $\hat g$ given in \eqref{ghat} is easily seen to be \begin{multline}
Z_{\hat g}(s,s') = \sum_{w\in W}\frac{(-1)^w}{\Delta(s)\Delta(s')} 
\int \prod_{i=1}^{a} d^{N-1}s^{(i)}  \times \\
\exp (2\pi \mathrm{i} s s^{(1)} + \pi \mathrm{i} k^{(1)} (s^{(1)})^2 + 2\pi \mathrm{i} s^{(1)}s^{(2)} 
+ \cdots + \pi \mathrm{i} k^{(a)} (s^{(a)})^2 + 2\pi \mathrm{i} s^{(a)} w(s'))
\end{multline}
which is \begin{equation}
= \sum_{w\in W}\frac{(-1)^w}{\Delta(s)\Delta(s')} 
\exp\left(  \frac{q}{p}\pi \mathrm{i} s^2  + \frac 1p 2\pi \mathrm{i} s w(s') + \frac{q'}{p} \pi \mathrm{i} (s')^2\right)\,.
\end{equation}
This is consistent with \eqref{eff}.

In a crude sense, our result means that the $SL(2,\bZ)$ transformation $\hat g$ \eqref{ghat}
given by the continued fraction \eqref{cfrac} behaves as a Chern--Simons coupling at a fractional level $q/p$.
This phenomenon has been noticed many times in the past.
For example, in the case of the $SL(2,\bZ)$ operation acting on 3d theories with $U(1)$ symmetries,
the theory corresponding to  $g\in SL(2,\bZ)$ is an Abelian Chern--Simons theory \cite{Witten:2003ya},
and this fractional effective Chern-Simons coupling explains why this Abelian Chern--Simons theory
can be used as a  low-energy description of the  fractional quantum Hall effect.

The same theory also appears in the description of D3-branes suspended between an NS 5-brane and a $(p,q)$ 5-brane,
whose study has a long history. 
The case of a D3-brane goes back at least to \cite{Kitao:1998mf}, see Sec.~3.2 and 3.3 there.
The full understanding of the multiple D3-brane case had to wait the seminal \cite{Gaiotto:2008ak}, 
see Sec.~8.2 and 8.3 there, where the issue of the fractional Chern--Simons levels was also mentioned in the Abelian case.
The same issue was also mentioned e.g.~in \cite[Sec.~2.2]{Gadde:2013sca} and \cite[Sec.~5.1.3]{Chung:2014qpa}.
Our analysis here can be thought of as giving a concrete meaning to the fractional Chern--Simons levels in the non-Abelian case.

\subsection{Torus bundles}
\label{sssec:torusbundle}

We can also consider the case of a circular quiver, 
obtained by adding a node with Chern--Simons level $k_0$ and 
closing the quiver chain of Figure \ref{fig:wall}.
The corresponding geometry is a torus bundle over $S^1$, where the torus 
is mapped back to itself with the $SL(2,\bZ)$ action 
\begin{equation}
h=T^{k^{(0)}}ST^{k^{(1)}}S\cdots T^{k^{(a)}}S
=T^{k^{(0)}}\hat g 
= T^{k^{(0)}} SgS.
\end{equation}
We expect the low energy 3d theory to only depend on conjugacy classes of $h$.

The superpotential is simply obtained by 
setting $\Phi=\Phi'$ and 
adding $- \frac{k_0}{2} \tr\Phi^2$ to \eqref{eff},
which is
\begin{equation}
  W_h = - \frac 12\Big( k^{(0)} - \frac{q}{p} - \frac{q'}{p} \Big) \tr\Phi^2 
  + \tr \mu_H^{(a)} \Phi + \tr \mu_C^{(0)} \Phi
  + \sum_{i=1}^{a}c^{(i)}\tr \mu_H^{(i-1)}\mu_C^{(i)} \,.
\end{equation}
Integrating out $\Phi$ leads to
\begin{equation}
  W_h = \frac 12\frac{p}{k^{(0)}p - q -q'}\tr (\mu_H^{(a)} + \mu_C^{(0)})^2
  + \sum_{i=1}^{a}c^{(i)}\tr \mu_H^{(i-1)}\mu_C^{(i)} \,.
\end{equation}
Now, we notice that $k^{(0)}p - q -q' = \tr h$. 
Rescaling the moment maps as 
\begin{equation}
  \mu_H^{(i)} \mapsto \Big(\prod_{j=0}^i c^{(i)}\Big)\tilde\mu_H^{(i)} \,, \quad
  \mu_C^{(i)}  \mapsto \Big(\prod_{j=0}^i c^{(i)}\Big)^{-1} \tilde\mu_C^{(i)} \,,
\end{equation} 
with $c^{(0)}:=p^{\frac 12}$ using the individual $J_{34}$ symmetry of the $i$-th copy of $T(SU(N))$ theory, we get
\begin{equation}
  W_h = \frac{1}{2\tr h} \tr (\mu_H^{(a)} + \mu_C^{(0)})^2
  + \sum_{i=1}^{a}\tr \mu_H^{(i-1)}\mu_C^{(i)} \,. \label{Wh}
\end{equation}
This final form clearly depends only on the conjugacy class of $h$, which was what we wanted to demonstrate.

We note here that we do not immediately see supersymmetry enhancement to \Nequals4 from this effective superpotential,
although the enhancement to \Nequals4 was found in the holographic dual in \cite{Assel:2018vtq} when $|\tr h| > 2$.
Let us now show field theoretically that it indeed has \Nequals4 supersymmetry, at least when $k^{(0)}$ is sufficiently large.
This is a refined version of an argument given in \cite{Gang:2018huc,Garozzo:2019ejm,Beratto:2020qyk}.\footnote{%
The argument in those reference went as follows. Before introducing $\Phi$ and gauging the last $SU(N)$, 
the duality wall theory has the chiral ring relation \eqref{crwall}, which becomes
$\tr (\mu_C^{(0)})^2 = \tr (\mu_H^{(a)})^2=0.$ 
Therefore, the superpotential $W_h$ above is equivalent to $ 
W_h' = \frac{1}{\tr h} \tr \mu_H^{(a)}  \mu_C^{(0)}
  + \sum_{i=1}^{a}\tr \mu_H^{(i-1)}\mu_C^{(i)},$
which preserves the $J_{34}$ symmetry. 
As we knew that this theory already has \Nequals3, this implies that it in fact has \Nequals4.
This argument is not very satisfactory, however. This is because the chiral ring relations $\tr (\mu_C^{(0)})^2 = \tr (\mu_H^{(a)})^2=0$  will get deformed under the superpotential  $W_h$ \eqref{Wh}. }

We start from the \Nequals4 duality wall theory for $h\in SL(2,\bZ)$,
and gauge the diagonal $SU(N)$ symmetry using \Nequals2 Chern--Simons multiplet at the level $k^{(0)}$,
without $\Phi$ and without any superpotential deformation. 
When $k^{(0)}$ is large enough, this theory is an \Nequals2 SCFT, preserving $J_{34}$.
Let us call this theory $T_0$.
The chiral ring relations 
\begin{equation}
\tr (\mu_C^{(0)})^2 = \tr (\mu_H^{(a)})^2=0. \label{naivecr}
\end{equation}
 hold in this theory.
This means that these operators are absent. 
The operator $\tr \mu_H^{(a)} \mu_C^{(0)}$ remains as a marginal chiral primary.

We now analyze the conformal manifold of this \Nequals2 theory $T_0$, applying the logic of \cite{Green:2010da}.
The possible  marginal deformation is $W= c \tr \mu_H^{(a)} \mu_C^{(0)}$,
which is uncharged under the flavor symmetry $J_{34}$.
Therefore, this deformation is exactly marginal, 
and the conformal manifold $\mathcal{M}$ close to $T_0$
is complex one dimensional and parametrized by $c$.
Furthermore, the $U(1)$  symmetry $J_{34}$ is preserved everywhere on  $\mathcal{M}$.
It is not clear how far in $c$ this conformal manifold $\mathcal{M}$  extends, however.
At least, when $k^{(0)}$ is sufficiently large, 
the \Nequals3 theory with the superpotential $W_h$ \eqref{Wh} 
is indeed close to the theory $T_0$.
As the \Nequals3 theory is   believed to be superconformal as is, 
it should be actually on $\mathcal{M}$ at some value of $c$, 
and  therefore the $J_{34}$ symmetry is unbroken there.
This implies that the supersymmetry enhances to \Nequals4.

\subsection{Some special cases}
\label{ssub:torus-bundles}

In this final subsection, we would like to mention some dualities, based on the fact that
some Seifert spaces can also be written as torus bundles; see \cite[Sec.~2.2]{HatcherOnline}. 
These can be viewed as $(T^2\times [0,1])/\sim$, with an identification $(t,0)\sim (g.t,1)$, where $t\in T^2$ and $h\in SL(2,\mathbb{Z})$. 
We just discussed its field theory realization in Sec.~\ref{sssec:torusbundle}.

The resulting theory only depends on the conjugacy class of $h$. Recall that an element is called hyperbolic, parabolic or elliptic depending on whether $|\mathrm{Tr}(h)|$ is $>2$, $=2$ or $<2$. A torus bundle can be Seifert only if $h$ is parabolic or elliptic.\footnote{Theories where $h$ is hyperbolic have appeared in \cite{Gang:2015wya,Assel:2018vtq}. The latter context provided a holographic dual, which confirmed ${\mathcal N}=4$ supersymmetry.} The parabolic case is well-known: when $h=T^k$, the space can also be written as an $S^1$-bundle over $T^2$ with $c_1=k$.

\begin{figure}
\centering
\begin{tikzpicture}
\node[hexagon] (TN) at (0,0) {$T_N$};
\node[circ,inner sep=4pt] (k1) at (2,1.2) {$2$};
\node[draw,rectangle] (n1) at (3,1.2) {$S$};
\node[circ,inner sep=4pt] (k11) at (4,1.2) {$2$};
\node[circ] (k2) at (2,0) {$-3$};
\node[circ] (k3) at (2,-1.2) {$-3$};
\draw (TN)--(k1)--(n1)--(k11);
\draw (TN)--(k2);
\draw (TN)--(k3);
\node (X) at (6,0){$\cong$};
\node[circ] (k4) at (8,0) {$-1$};
\node[draw,rectangle] (n2) at (10,0) {$S$};
\draw (k4) edge[bend left] (n2);
\draw (k4) edge[bend right] (n2);
\end{tikzpicture}
\caption{A duality suggested by two equivalent descriptions of a space as a Seifert manifold or as a torus bundle. Recall that in our notation the numbers in the nodes are the CS levels. \label{fig:duality}}
\end{figure}
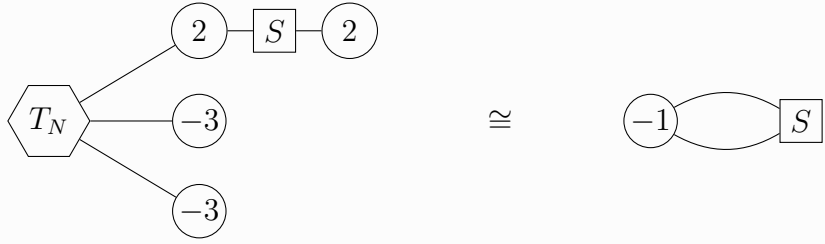

The elliptic case is perhaps more interesting. There are a few cases: 
\begin{itemize}
	\item The Seifert $(0;1/2,1/2,-1/2,-1/2)$ is the $T^2$-bundle with $h=-1_2$.
	\item The Seifert $(0;1/2,-1/4,-1/4)$ is the $T^2$-bundle with $h=S$; this acts as a $\pi/2$ rotation on the torus with $\tau=\mathrm{i}$.
	\item $(0;1/2,-1/3,-1/6)$ corresponds to the $T^2$-bundle with $h=ST$, acting as a $\pi/6$ rotation on the torus with $\tau=\mathrm{e}^{\pi \mathrm{i}/3}$.
	\item $(0;2/3,-1/3,-1/3)$ corresponds to the $T^2$-bundle with $h=T^{-1}S= (ST)^2$, acting as a $\pi/3$ rotation on the torus with $\tau=\mathrm{e}^{\pi \mathrm{i}/3}$.
\end{itemize}
(Notice that $S$, $ST$, $(ST)^2$ and their inverses represent all the elliptic conjugacy classes.) These identifications suggest dualities between the general Seifert theory of Fig.~\ref{fig:theory} and the torus bundle theory discussed in Sec.~\ref{sssec:torusbundle}. 
For example, the last identification in the list above suggests the duality in Fig.~\ref{fig:duality}. It would be interesting to check these dualities by computing protected quantities on both sides.\footnote{The $S^3$ partition function appears to be ill-defined; on the torus bundle side, this can be checked by applying the formulas in \cite[App.~C]{Assel:2018vtq}. The superconformal index appears to be more promising; see \cite{Garozzo:2019ejm,Beratto:2020qyk} for this computation in similar theories.} 
We note that $\sum 1/k_i=0$ for all the Seifert spaces listed above.
This means that they are all \Nequals4, agreeing with the fact that the supposedly dual $T^2$-bundle theories are also \Nequals4, as we just saw in Sec.~\ref{sssec:torusbundle}.

Finally, notice a certain similarity to the way ${\mathcal N}=3$ theories were defined in four dimensions \cite{Garcia-Etxebarria:2015wns}: there, a quotient was taken of ${\mathcal N}=4$ super-Yang--Mills by the elements in the above list, with $\tau$ fixed to the corresponding values. In general, the torus bundle theories can also be thought of as ${\mathcal N}=4$, $d=4$ super-Yang--Mills on a circle with a $g\in SL(2,\mathbb{Z})$ monodromy acting on $\tau$. This seems to suggest a compactification relation between the two sets of theories.

\def\arxivfont{\rm}
\bibliographystyle{ytamsalpha}

\baselineskip=.98\baselineskip

\bibliography{ref}

\newcommand{\etalchar}[1]{$^{#1}$}
\providecommand{\bysame}{\leavevmode\hbox to3em{\hrulefill}\thinspace}
\providecommand{\MR}{\relax\ifhmode\unskip\space\fi MR }
\providecommand{\MRhref}[2]{%
  \href{http://www.ams.org/mathscinet-getitem?mr=#1}{#2}
}
\providecommand{\href}[2]{#2}
\providecommand{\doihref}[2]{\href{#1}{#2}}
\providecommand{\arxivfont}{\tt}
\begin{thebibliography}{GLMMS19}

\bibitem[ABJ08]{Aharony:2008gk}
O.~Aharony, O.~Bergman, and D.~L. Jafferis, \emph{{Fractional M2-Branes}},
  \doihref{http://dx.doi.org/10.1088/1126-6708/2008/11/043}{JHEP \textbf{11}
  (2008) 043}, \href{http://arxiv.org/abs/0807.4924}{{\arxivfont
  arXiv:0807.4924 [hep-th]}}.

\bibitem[ABJM08]{Aharony:2008ug}
O.~Aharony, O.~Bergman, D.~L. Jafferis, and J.~Maldacena,
  \emph{{${\mathcal{N}}\!=6$ Superconformal Chern-Simons-Matter Theories,
  M2-Branes and Their Gravity Duals}},
  \doihref{http://dx.doi.org/10.1088/1126-6708/2008/10/091}{JHEP \textbf{10}
  (2008) 091}, \href{http://arxiv.org/abs/0806.1218}{{\arxivfont
  arXiv:0806.1218 [hep-th]}}.

\bibitem[ASNW16]{Assel:2016lad}
B.~Assel, S.~Sch{\"a}fer-Nameki, and J.-M. Wong, \emph{{M5-branes on
  $S^{2}\times M_{4}$: Nahm\textquoteright{}s equations and 4d topological
  sigma-models}}, \doihref{http://dx.doi.org/10.1007/JHEP09(2016)120}{JHEP
  \textbf{09} (2016) 120}, \href{http://arxiv.org/abs/1604.03606}{{\arxivfont
  arXiv:1604.03606 [hep-th]}}.

\bibitem[AT18]{Assel:2018vtq}
B.~Assel and A.~Tomasiello, \emph{{Holographic duals of 3d S-fold CFTs}},
  \doihref{http://dx.doi.org/10.1007/JHEP06(2018)019}{JHEP \textbf{06} (2018)
  019}, \href{http://arxiv.org/abs/1804.06419}{{\arxivfont arXiv:1804.06419
  [hep-th]}}.

\bibitem[BK10]{Bashkirov:2010kz}
D.~Bashkirov and A.~Kapustin, \emph{{Supersymmetry Enhancement by Monopole
  Operators}}, \doihref{http://dx.doi.org/10.1007/JHEP05(2011)015}{JHEP
  \textbf{05} (2011) 015}, \href{http://arxiv.org/abs/1007.4861}{{\arxivfont
  arXiv:1007.4861 [hep-th]}}.

\bibitem[BL07]{Bagger:2007jr}
J.~Bagger and N.~Lambert, \emph{{Gauge Symmetry and Supersymmetry of Multiple
  M2-Branes}}, \doihref{http://dx.doi.org/10.1103/PhysRevD.77.065008}{Phys.
  Rev. D \textbf{77} (2008) 065008},
  \href{http://arxiv.org/abs/0711.0955}{{\arxivfont arXiv:0711.0955 [hep-th]}}.

\bibitem[BLS08]{Bandres:2008vf}
M.~A. Bandres, A.~E. Lipstein, and J.~H. Schwarz, \emph{{${\mathcal{N}}\!=8$
  Superconformal Chern-Simons Theories}},
  \doihref{http://dx.doi.org/10.1088/1126-6708/2008/05/025}{JHEP \textbf{05}
  (2008) 025}, \href{http://arxiv.org/abs/0803.3242}{{\arxivfont
  arXiv:0803.3242 [hep-th]}}.

\bibitem[BMS20]{Beratto:2020qyk}
E.~Beratto, N.~Mekareeya, and M.~Sacchi, \emph{{Marginal operators and
  supersymmetry enhancement in 3d $S$-fold SCFTs}},
  \doihref{http://dx.doi.org/10.1007/JHEP12(2020)017}{JHEP \textbf{12} (2020)
  017}, \href{http://arxiv.org/abs/2009.10123}{{\arxivfont arXiv:2009.10123
  [hep-th]}}.

\bibitem[Boc09]{bock}
C.~Bock, \emph{On low-dimensional solvmanifolds},
  \doihref{http://dx.doi.org/10.4310/AJM.2016.v20.n2.a1}{Asian Journal of
  Mathematics \textbf{20} (2016) 199--262},
  \href{http://arxiv.org/abs/0903.2926}{{\arxivfont arXiv:0903.2926
  [math.DG]}}.

\bibitem[BP11]{Benvenuti:2011ga}
S.~Benvenuti and S.~Pasquetti, \emph{{3D-Partition Functions on the Sphere:
  Exact Evaluation and Mirror Symmetry}},
  \doihref{http://dx.doi.org/10.1007/JHEP05(2012)099}{JHEP \textbf{05} (2012)
  099}, \href{http://arxiv.org/abs/1105.2551}{{\arxivfont arXiv:1105.2551
  [hep-th]}}.

\bibitem[Bru96]{brunella}
M.~Brunella, \emph{On transversely holomorphic flows {I}},
  \doihref{http://dx.doi.org/10.1007/s002220050098}{Inventiones mathematicae
  \textbf{126} (1996) 265--279}.

\bibitem[BS03]{BuyaloSvetlov}
S.~V. Buyalo and P.~V. Svetlov, \emph{Topological and geometric properties of
  graph manifolds},
  \doihref{http://dx.doi.org/10.1090/S1061-0022-05-00852-6}{St. Petersburg
  Math. J. \textbf{16} (2005) 297--340},
  \href{http://arxiv.org/abs/math.GT/0309192}{{\arxivfont
  arXiv:math.GT/0309192}}.

\bibitem[BSVP99]{Bergshoeff:1999db}
E.~Bergshoeff, E.~Sezgin, and A.~Van~Proeyen, \emph{{(2,0) tensor multiplets
  and conformal supergravity in D = 6}},
  \doihref{http://dx.doi.org/10.1088/0264-9381/16/10/311}{Class. Quant. Grav.
  \textbf{16} (1999) 3193--3206},
  \href{http://arxiv.org/abs/hep-th/9904085}{{\arxivfont
  arXiv:hep-th/9904085}}.

\bibitem[BTX10]{Benini:2010uu}
F.~Benini, Y.~Tachikawa, and D.~Xie, \emph{{Mirrors of 3D Sicilian Theories}},
  \doihref{http://dx.doi.org/10.1007/JHEP09(2010)063}{JHEP \textbf{09} (2010)
  063}, \href{http://arxiv.org/abs/1007.0992}{{\arxivfont arXiv:1007.0992
  [hep-th]}}.

\bibitem[CDF{\etalchar{+}}12a]{Closset:2012vg}
C.~Closset, T.~T. Dumitrescu, G.~Festuccia, Z.~Komargodski, and N.~Seiberg,
  \emph{{Contact Terms, Unitarity, and F-Maximization in Three-Dimensional
  Superconformal Theories}},
  \doihref{http://dx.doi.org/10.1007/JHEP10(2012)053}{JHEP \textbf{10} (2012)
  053}, \href{http://arxiv.org/abs/1205.4142}{{\arxivfont arXiv:1205.4142
  [hep-th]}}.

\bibitem[CDF{\etalchar{+}}12b]{Closset:2012vp}
C.~Closset, T.~T. Dumitrescu, G.~Festuccia, Z.~Komargodski, and N.~Seiberg,
  \emph{{Comments on Chern-Simons Contact Terms in Three Dimensions}},
  \doihref{http://dx.doi.org/10.1007/JHEP09(2012)091}{JHEP \textbf{09} (2012)
  091}, \href{http://arxiv.org/abs/1206.5218}{{\arxivfont arXiv:1206.5218
  [hep-th]}}.

\bibitem[CDFK12]{Closset:2012ru}
C.~Closset, T.~T. Dumitrescu, G.~Festuccia, and Z.~Komargodski,
  \emph{{Supersymmetric Field Theories on Three-Manifolds}},
  \doihref{http://dx.doi.org/10.1007/JHEP05(2013)017}{JHEP \textbf{05} (2013)
  017}, \href{http://arxiv.org/abs/1212.3388}{{\arxivfont arXiv:1212.3388
  [hep-th]}}.

\bibitem[CDGS14]{Chung:2014qpa}
H.-J. Chung, T.~Dimofte, S.~Gukov, and P.~Su\l{}kowski, \emph{{3d-3d
  Correspondence Revisited}},
  \doihref{http://dx.doi.org/10.1007/JHEP04(2016)140}{JHEP \textbf{04} (2016)
  140}, \href{http://arxiv.org/abs/1405.3663}{{\arxivfont arXiv:1405.3663
  [hep-th]}}.

\bibitem[CGK20]{Cho:2020ljj}
G.~Y. Cho, D.~Gang, and H.-C. Kim, \emph{{M-Theoretic Genesis of Topological
  Phases}}, \doihref{http://dx.doi.org/10.1007/JHEP11(2020)115}{JHEP
  \textbf{11} (2020) 115}, \href{http://arxiv.org/abs/2007.01532}{{\arxivfont
  arXiv:2007.01532 [hep-th]}}.

\bibitem[CGK22]{Choi:2022dju}
S.~Choi, D.~Gang, and H.-C. Kim, \emph{{Infrared Phases of 3D Class R
  Theories}}, \href{http://arxiv.org/abs/2206.11982}{{\arxivfont
  arXiv:2206.11982 [hep-th]}}.

\bibitem[DGG11a]{Dimofte:2011ju}
T.~Dimofte, D.~Gaiotto, and S.~Gukov, \emph{{Gauge Theories Labelled by
  Three-Manifolds}},
  \doihref{http://dx.doi.org/10.1007/s00220-013-1863-2}{Commun. Math. Phys.
  \textbf{325} (2014) 367--419},
  \href{http://arxiv.org/abs/1108.4389}{{\arxivfont arXiv:1108.4389 [hep-th]}}.

\bibitem[DGG11b]{Dimofte:2011py}
T.~Dimofte, D.~Gaiotto, and S.~Gukov, \emph{{3-Manifolds and 3D Indices}},
  \doihref{http://dx.doi.org/10.4310/ATMP.2013.v17.n5.a3}{Adv. Theor. Math.
  Phys. \textbf{17} (2013) 975--1076},
  \href{http://arxiv.org/abs/1112.5179}{{\arxivfont arXiv:1112.5179 [hep-th]}}.

\bibitem[DGG13]{Dimofte:2013iv}
T.~Dimofte, M.~Gabella, and A.~B. Goncharov, \emph{{K-Decompositions and 3D
  Gauge Theories}}, \doihref{http://dx.doi.org/10.1007/JHEP11(2016)151}{JHEP
  \textbf{11} (2016) 151}, \href{http://arxiv.org/abs/1301.0192}{{\arxivfont
  arXiv:1301.0192 [hep-th]}}.

\bibitem[EKSNW19]{Eckhard:2019jgg}
J.~Eckhard, H.~Kim, S.~Schäfer-Nameki, and B.~Willett, \emph{{Higher-Form
  Symmetries, Bethe Vacua, and the 3D-3D Correspondence}},
  \doihref{http://dx.doi.org/10.1007/JHEP01(2020)101}{JHEP \textbf{01} (2020)
  101}, \href{http://arxiv.org/abs/1910.14086}{{\arxivfont arXiv:1910.14086
  [hep-th]}}.

\bibitem[FS11]{Festuccia:2011ws}
G.~Festuccia and N.~Seiberg, \emph{{Rigid Supersymmetric Theories in Curved
  Superspace}}, \doihref{http://dx.doi.org/10.1007/JHEP06(2011)114}{JHEP
  \textbf{06} (2011) 114}, \href{http://arxiv.org/abs/1105.0689}{{\arxivfont
  arXiv:1105.0689 [hep-th]}}.

\bibitem[GER15]{Garcia-Etxebarria:2015wns}
I.~Garc\'\i{}a-Etxebarria and D.~Regalado, \emph{{$ \mathcal{N}=3 $ four
  dimensional field theories}},
  \doihref{http://dx.doi.org/10.1007/JHEP03(2016)083}{JHEP \textbf{03} (2016)
  083}, \href{http://arxiv.org/abs/1512.06434}{{\arxivfont arXiv:1512.06434
  [hep-th]}}.

\bibitem[GGP13]{Gadde:2013sca}
A.~Gadde, S.~Gukov, and P.~Putrov, \emph{{Fivebranes and 4-Manifolds}},
  \doihref{http://dx.doi.org/10.1007/978-3-319-43648-7_7}{Prog. Math.
  \textbf{319} (2016) 155--245},
  \href{http://arxiv.org/abs/1306.4320}{{\arxivfont arXiv:1306.4320 [hep-th]}}.

\bibitem[Ghy96]{ghys}
{\'E}.~Ghys, \emph{On transversely holomorphic flows {II}},
  \doihref{http://dx.doi.org/10.1007/s002220050099}{Inventiones mathematicae
  \textbf{126} (1996) 281--286}.

\bibitem[GKRY15]{Gang:2015wya}
D.~Gang, N.~Kim, M.~Romo, and M.~Yamazaki, \emph{{Aspects of Defects in 3d-3d
  Correspondence}}, \doihref{http://dx.doi.org/10.1007/JHEP10(2016)062}{JHEP
  \textbf{10} (2016) 062}, \href{http://arxiv.org/abs/1510.05011}{{\arxivfont
  arXiv:1510.05011 [hep-th]}}.

\bibitem[GKS{\etalchar{+}}10]{Green:2010da}
D.~Green, Z.~Komargodski, N.~Seiberg, Y.~Tachikawa, and B.~Wecht,
  \emph{{Exactly Marginal Deformations and Global Symmetries}},
  \doihref{http://dx.doi.org/10.1007/JHEP06(2010)106}{JHEP \textbf{06} (2010)
  106}, \href{http://arxiv.org/abs/1005.3546}{{\arxivfont arXiv:1005.3546
  [hep-th]}}.

\bibitem[GLMMS19]{Garozzo:2019ejm}
I.~Garozzo, G.~Lo~Monaco, N.~Mekareeya, and M.~Sacchi, \emph{{Supersymmetric
  Indices of 3d $S$-fold SCFTs}},
  \doihref{http://dx.doi.org/10.1007/JHEP08(2019)008}{JHEP \textbf{08} (2019)
  008}, \href{http://arxiv.org/abs/1905.07183}{{\arxivfont arXiv:1905.07183
  [hep-th]}}.

\bibitem[GPPV17]{Gukov:2017kmk}
S.~Gukov, D.~Pei, P.~Putrov, and C.~Vafa, \emph{{BPS Spectra and 3-Manifold
  Invariants}}, \doihref{http://dx.doi.org/10.1142/S0218216520400039}{J. Knot
  Theor. Ramifications \textbf{29} (2020) 2040003},
  \href{http://arxiv.org/abs/1701.06567}{{\arxivfont arXiv:1701.06567
  [hep-th]}}.

\bibitem[GPV16]{Gukov:2016gkn}
S.~Gukov, P.~Putrov, and C.~Vafa, \emph{{Fivebranes and 3-Manifold Homology}},
  \doihref{http://dx.doi.org/10.1007/JHEP07(2017)071}{JHEP \textbf{07} (2017)
  071}, \href{http://arxiv.org/abs/1602.05302}{{\arxivfont arXiv:1602.05302
  [hep-th]}}.

\bibitem[Gus07]{Gustavsson:2007vu}
A.~Gustavsson, \emph{{Algebraic Structures on Parallel M2-Branes}},
  \doihref{http://dx.doi.org/10.1016/j.nuclphysb.2008.11.014}{Nucl. Phys. B
  \textbf{811} (2009) 66--76},
  \href{http://arxiv.org/abs/0709.1260}{{\arxivfont arXiv:0709.1260 [hep-th]}}.

\bibitem[GW08a]{Gaiotto:2008sd}
D.~Gaiotto and E.~Witten, \emph{{Janus Configurations, Chern-Simons Couplings,
  and the Theta-Angle in ${\mathcal{N}}\!=4$ Super Yang-Mills Theory}},
  \doihref{http://dx.doi.org/10.1007/JHEP06(2010)097}{JHEP \textbf{06} (2010)
  097}, \href{http://arxiv.org/abs/0804.2907}{{\arxivfont arXiv:0804.2907
  [hep-th]}}.

\bibitem[GW08b]{Gaiotto:2008ak}
D.~Gaiotto and E.~Witten, \emph{{S-Duality of Boundary Conditions in
  ${\mathcal{N}}\!=4$ Super Yang-Mills Theory}},
  \doihref{http://dx.doi.org/10.4310/ATMP.2009.v13.n3.a5}{Adv. Theor. Math.
  Phys. \textbf{13} (2009) 721--896},
  \href{http://arxiv.org/abs/0807.3720}{{\arxivfont arXiv:0807.3720 [hep-th]}}.

\bibitem[GY07]{Gaiotto:2007qi}
D.~Gaiotto and X.~Yin, \emph{{Notes on Superconformal Chern-Simons-Matter
  Theories}}, \doihref{http://dx.doi.org/10.1088/1126-6708/2007/08/056}{JHEP
  \textbf{08} (2007) 056}, \href{http://arxiv.org/abs/0704.3740}{{\arxivfont
  arXiv:0704.3740 [hep-th]}}.

\bibitem[GY18a]{Gang:2018wek}
D.~Gang and K.~Yonekura, \emph{{Symmetry Enhancement and Closing of Knots in
  3D/3D Correspondence}},
  \doihref{http://dx.doi.org/10.1007/JHEP07(2018)145}{JHEP \textbf{07} (2018)
  145}, \href{http://arxiv.org/abs/1803.04009}{{\arxivfont arXiv:1803.04009
  [hep-th]}}.

\bibitem[GY18b]{Gang:2018huc}
D.~Gang and M.~Yamazaki, \emph{{Three-Dimensional Gauge Theories with
  Supersymmetry Enhancement}},
  \doihref{http://dx.doi.org/10.1103/PhysRevD.98.121701}{Phys. Rev. D
  \textbf{98} (2018) 121701},
  \href{http://arxiv.org/abs/1806.07714}{{\arxivfont arXiv:1806.07714
  [hep-th]}}.

\bibitem[Ham95]{Hamilton}
R.~S. Hamilton, \emph{The formation of singularities in the {R}icci flow},
  \doihref{http://dx.doi.org/10.4310/SDG.1993.v2.n1.a2}{Surveys in differential
  geometry \textbf{2} (1995) 7--136}. Proceedings of the conference on geometry
  and topology held at Harvard University, April 23-25, 1993.

\bibitem[Hat07]{HatcherOnline}
A.~Hatcher, \emph{Notes on basic 3-manifold topology}.
  \url{https://pi.math.cornell.edu/~hatcher/3M/3Mdownloads.html}.

\bibitem[HLL{\etalchar{+}}08a]{Hosomichi:2008jd}
K.~Hosomichi, K.-M. Lee, S.~Lee, S.~Lee, and J.~Park,
  \emph{{${\mathcal{N}}{=}4$ Superconformal Chern-Simons Theories with Hyper
  and Twisted Hyper Multiplets}},
  \doihref{http://dx.doi.org/10.1088/1126-6708/2008/07/091}{JHEP \textbf{07}
  (2008) 091}, \href{http://arxiv.org/abs/0805.3662}{{\arxivfont
  arXiv:0805.3662 [hep-th]}}.

\bibitem[HLL{\etalchar{+}}08b]{Hosomichi:2008jb}
K.~Hosomichi, K.-M. Lee, S.~Lee, S.~Lee, and J.~Park,
  \emph{{${\mathcal{N}}{=}5,6$ Superconformal Chern-Simons Theories and
  M2-Branes on Orbifolds}},
  \doihref{http://dx.doi.org/10.1088/1126-6708/2008/09/002}{JHEP \textbf{09}
  (2008) 002}, \href{http://arxiv.org/abs/0806.4977}{{\arxivfont
  arXiv:0806.4977 [hep-th]}}.

\bibitem[Joh79]{J}
K.~Johannson, \doihref{http://dx.doi.org/10.1007/BFb0085406}{\emph{Homotopy
  equivalences of {$3$}-manifolds with boundaries}}, Lecture Notes in
  Mathematics, vol. 761, Springer, Berlin, 1979.

\bibitem[JS79]{JS}
W.~H. Jaco and P.~B. Shalen, \emph{Seifert fibered spaces in {$3$}-manifolds},
  \href{https://doi.org/10.1090/memo/0220}{Mem. Amer. Math. Soc. \textbf{21}
  (1979) viii+192}.

\bibitem[Kac77]{Kac}
V.~G. Kac, \emph{Lie superalgebras},
  \doihref{http://dx.doi.org/10.1016/0001-8708(77)90017-2}{Advances in Math.
  \textbf{26} (1977) 8--96}. Dedicated to the Memory of My Father, G. M. Kac
  (September 15, 1918 -- September 25, 1974).

\bibitem[KOO98]{Kitao:1998mf}
T.~Kitao, K.~Ohta, and N.~Ohta, \emph{{Three-Dimensional Gauge Dynamics from
  Brane Configurations with $(p,q)$-Fivebrane}},
  \doihref{http://dx.doi.org/10.1016/S0550-3213(98)00726-3}{Nucl. Phys. B
  \textbf{539} (1999) 79--106},
  \href{http://arxiv.org/abs/hep-th/9808111}{{\arxivfont
  arXiv:hep-th/9808111}}.

\bibitem[KS09]{Kapustin:2009cd}
A.~Kapustin and N.~Saulina, \emph{{Chern-Simons-Rozansky-Witten Topological
  Field Theory}},
  \doihref{http://dx.doi.org/10.1016/j.nuclphysb.2009.07.006}{Nucl. Phys. B
  \textbf{823} (2009) 403--427},
  \href{http://arxiv.org/abs/0904.1447}{{\arxivfont arXiv:0904.1447 [hep-th]}}.

\bibitem[KTZ12]{Klare:2012gn}
C.~Klare, A.~Tomasiello, and A.~Zaffaroni, \emph{{Supersymmetry on Curved
  Spaces and Holography}},
  \doihref{http://dx.doi.org/10.1007/JHEP08(2012)061}{JHEP \textbf{08} (2012)
  061}, \href{http://arxiv.org/abs/1205.1062}{{\arxivfont arXiv:1205.1062
  [hep-th]}}.

\bibitem[Mar16]{martelli-book}
B.~Martelli, \emph{An introduction to geometric topology},
  \href{http://arxiv.org/abs/1610.02592}{{\arxivfont arXiv:1610.02592
  [math.GT]}}.
  \url{https://people.dm.unipi.it/martelli/geometric_topology.html}.

\bibitem[Neu97]{Neumann}
W.~D. Neumann, \emph{Commensurability and virtual fibration for graph
  manifolds}, \doihref{http://dx.doi.org/10.1016/0040-9383(96)00014-6}{Topology
  \textbf{36} (1997) 355--378}.

\bibitem[NTY11]{Nishioka:2011dq}
T.~Nishioka, Y.~Tachikawa, and M.~Yamazaki, \emph{{3D Partition Function as
  Overlap of Wavefunctions}},
  \doihref{http://dx.doi.org/10.1007/JHEP08(2011)003}{JHEP \textbf{08} (2011)
  003}, \href{http://arxiv.org/abs/1105.4390}{{\arxivfont arXiv:1105.4390
  [hep-th]}}.

\bibitem[Per02]{Perelman}
G.~Perelman, \emph{The entropy formula for the {Ricci} flow and its geometric
  applications}, \href{http://arxiv.org/abs/math.DG/0211159}{{\arxivfont
  arXiv:math.DG/0211159}}.

\bibitem[PY15]{Pei:2015jsa}
D.~Pei and K.~Ye, \emph{{A 3D-3D Appetizer}},
  \doihref{http://dx.doi.org/10.1007/JHEP11(2016)008}{JHEP \textbf{11} (2016)
  008}, \href{http://arxiv.org/abs/1503.04809}{{\arxivfont arXiv:1503.04809
  [hep-th]}}.

\bibitem[Sch04]{Schwarz:2004yj}
J.~H. Schwarz, \emph{{Superconformal Chern-Simons Theories}},
  \doihref{http://dx.doi.org/10.1088/1126-6708/2004/11/078}{JHEP \textbf{11}
  (2004) 078}, \href{http://arxiv.org/abs/hep-th/0411077}{{\arxivfont
  arXiv:hep-th/0411077}}.

\bibitem[Sco83]{Scott}
P.~Scott, \emph{The geometries of {$3$}-manifolds},
  \href{https://doi.org/10.1112/blms/15.5.401}{Bull. London Math. Soc.
  \textbf{15} (1983) 401--487}.

\bibitem[SST12]{Samtleben:2012ua}
H.~Samtleben, E.~Sezgin, and D.~Tsimpis, \emph{{Rigid 6D supersymmetry and
  localization}}, \doihref{http://dx.doi.org/10.1007/JHEP03(2013)137}{JHEP
  \textbf{03} (2013) 137}, \href{http://arxiv.org/abs/1212.4706}{{\arxivfont
  arXiv:1212.4706 [hep-th]}}.

\bibitem[ST08]{Schnabl:2008wj}
M.~Schnabl and Y.~Tachikawa, \emph{{Classification of ${\mathcal{N}}\!=6$
  Superconformal Theories of Abjm Type}},
  \doihref{http://dx.doi.org/10.1007/JHEP09(2010)103}{JHEP \textbf{09} (2010)
  103}, \href{http://arxiv.org/abs/0807.1102}{{\arxivfont arXiv:0807.1102
  [hep-th]}}.

\bibitem[Tac15]{Tachikawa:2015bga}
Y.~Tachikawa, \emph{{A review of the $T_N$ theory and its cousins}},
  \doihref{http://dx.doi.org/10.1093/ptep/ptv098}{PTEP \textbf{2015} (2015)
  11B102}, \href{http://arxiv.org/abs/1504.01481}{{\arxivfont arXiv:1504.01481
  [hep-th]}}.

\bibitem[Thu82]{Thurston}
W.~P. Thurston, \emph{Three-dimensional manifolds, {K}leinian groups and
  hyperbolic geometry},
  \href{https://doi.org/10.1090/S0273-0979-1982-15003-0}{Bull. Amer. Math. Soc.
  (N.S.) \textbf{6} (1982) 357--381}.

\bibitem[TY11]{Terashima:2011qi}
Y.~Terashima and M.~Yamazaki, \emph{{$SL(2,R)$ Chern-Simons, Liouville, and
  Gauge Theory on Duality Walls}},
  \doihref{http://dx.doi.org/10.1007/JHEP08(2011)135}{JHEP \textbf{08} (2011)
  135}, \href{http://arxiv.org/abs/1103.5748}{{\arxivfont arXiv:1103.5748
  [hep-th]}}.

\bibitem[Wit03]{Witten:2003ya}
E.~Witten, \emph{{$SL(2,Z)$ Action on Three-Dimensional Conformal Field
  Theories with Abelian Symmetry}}, {From Fields to Strings: Circumnavigating
  Theoretical Physics: a Conference in Tribute to Ian Kogan}, 7 2003,
  pp.~1173--1200. \href{http://arxiv.org/abs/hep-th/0307041}{{\arxivfont
  arXiv:hep-th/0307041}}.

\bibitem[WJ86]{welsh}
D.~J. Welsh~Jr, \emph{Manifolds that admit parallel vector fields}, Illinois
  Journal of Mathematics \textbf{30} (1986) 9--18.

\end{thebibliography}

\end{document}